\documentclass[twocolumn]{aastex63}

\usepackage{amsmath}
\usepackage{graphicx}
\usepackage{natbib}
\usepackage[caption=false]{subfig}
\usepackage[nameinlink]{cleveref}
\usepackage{verbatim}
\usepackage[caption=false]{subfig}

\newcommand\gaia{\textit{Gaia}\ }

\received{July 28, 2020}
\revised{October 20, 2020}
\accepted{October 20, 2020}
\submitjournal{ApJ}

\shorttitle{Galactic Warp as Revealed by Asymmetries in Galactic Disk Kinematics}
\shortauthors{Cheng et al.}
\graphicspath{{./}{figures/}}

\begin{document}

\title{Exploring the Galactic Warp Through Asymmetries in the Kinematics of the Galactic Disk}

\correspondingauthor{Xinlun Cheng}
\email{xc7ts@virginia.edu}

\author[0000-0002-7009-3957]{Xinlun Cheng}
\affiliation{Department of Astronomy, University of Virginia, Charlottesville, VA 22904-4325, USA}

\author[0000-0001-5261-4336]{Borja Anguiano}
\affiliation{Department of Astronomy, University of Virginia, Charlottesville, VA 22904-4325, USA}

\author[0000-0003-2025-3147]{Steven R. Majewski}
\affiliation{Department of Astronomy, University of Virginia, Charlottesville, VA 22904-4325, USA}

\author[0000-0003-2969-2445]{Christian Hayes}
\affiliation{Department of Astronomy, University of Virginia, Charlottesville, VA 22904-4325, USA}

\author{Phil Arras}
\affiliation{Department of Astronomy, University of Virginia, Charlottesville, VA 22904-4325, USA}

\author{Cristina Chiappini}
\affiliation{Leibniz-Institut fur Astrophysik Potsdam (AIP), An der Sternwarte 16, D-14482 Potsdam, Germany}

\author{Sten Hasselquist}
\affiliation{Department of Physics and Astronomy, University of Utah, 115 S. 1400 E., Salt Lake City, UT 84112, USA}

\author{Anna Bárbara de Andrade Queiroz}
\affiliation{Leibniz-Institut fur Astrophysik Potsdam (AIP), An der Sternwarte 16, D-14482 Potsdam, Germany}

\author{Christian Nitschelm}
\affiliation{Centro de Astronom{\'i}a (CITEVA), Universidad de Antofagasta, Avenida Angamos 601, Antofagasta 1270300, Chile}

\author{Domingo Aníbal García-Hernández}
\affiliation{Instituto de Astrofısica de Canarias, 38205 La Laguna, Tenerife, Spain}

\author{Richard R. Lane}
\affiliation{Instituto de Astrofısica, Pontificia Universidad Católica de Chile, Av. Vicuna Mackenna 4860, 782-0436 Macul, Santiago, Chile}
\affiliation{Instituto de Astronom\'ia y Ciencias Planetarias, Universidad de Atacama, Copayapu 485, Copiap\'o, Chile}

\author{Alexandre Roman-Lopes}
\affiliation{Departamento de F\'isica, Facultad de Ciencias, Universidad de La Serena, Cisternas 1200, La Serena, Chile}

\author{Peter M. Frinchaboy}
\affiliation{Department of Physics \& Astronomy, Texas Christian University, Fort Worth, TX 76129, USA}

\begin{abstract}
Previous analyses of large databases of Milky Way stars have revealed the stellar disk of our Galaxy to be warped and that this imparts a strong signature on the kinematics of stars beyond the solar neighborhood. However, due to the limitation of accurate distance estimates, many attempts to explore the extent of these Galactic features have generally been restricted to a volume near the Sun. By combining Gaia DR2 astrometric solution, StarHorse distance and stellar abundances from the APOGEE survey, we present the most detailed and radially expansive study yet of the vertical and radial motions of stars in the Galactic disk. We map stellar velocity with respect to their Galactocentric radius, angular momentum, and azimuthal angle and assess their relation to the warp. A decrease in vertical velocity is discovered at Galactocentric radius $R=13\ \text{kpc}$ and angular momentum $L_z=2800\ \text{kpc}\ \text{km}\ \text{s}^{-1}$. Smaller ripples in vertical and radial velocity are also discovered superposed on the main trend. We also discovered that trends in the vertical velocity with azimuthal angle are not symmetric about the peak, suggesting the warp to be lopsided. To explain the global trend in vertical velocity, we built a simple analytical model of the Galactic warp. Our best fit yields a starting radius of $8.87^{+0.08}_{-0.09}\ \text{kpc}$ and precession rate of $13.57^{+0.20}_{-0.18}\ \text{km}\ \text{s}^{-1}\ \text{kpc}^{-1}$. These parameters remain consistent across stellar age groups, a result that supports the notion that the warp is the result of an external, gravitationally induced phenomenon.
\end{abstract}


\section{Introduction} \label{sec:intro}

Disk warps are common features of spiral galaxies \citep{Bosma1978,Binney1992}, and the presence of a warp in the outer Milky Way disk has been long-established, as seen in its H I \citep[e.g.,][]{Kerr1957,Westerhout1957,Weaver1974,Levine2006,Voskes2006}, dust \citep{Freudenreich1994}, star-forming regions \citep{Wouterloot1990} and stellar disk components \citep[e.g.,][and references therein]{Amores2017}. The ubiquity of warps suggests that they are either repeatedly regenerated or long-lived phenomena in the lives of galaxy disks \citep{Sellwood2013}.

While the origin of the Galactic warp still invites controversy, the fact that the stellar warp follows the same topology as the gaseous one is evidence that the warp is gravitationally induced \citep[e.g.,][]{Miyamoto1988,Drimmel2000}. Interactions with massive satellite galaxies can also affect the outskirts of galaxy disks,  where the most likely candidates to create a warped outer disk in the Milky Way are the Sagittarius (Sgr) dwarf spheroidal (dSph) galaxy \citep{Ibata1998,Laporte2019}, and the Magellanic Clouds \citep{Weinberg&Blitz2006,Garavito2019}. External torques on galaxy disks have also been identified with the accretion of intergalactic matter \citep{Ostriker1989, Wang2020}, intergalactic magnetic fields \citep{Battaner1990,Guijarro2010} and mis-aligned dark halos \citep{Sparke1988, Widrow2014, Amores2017}. Moreover, disk instability has also been attributed to the cause of the warp. For instance, \citet{Chen2019} probed line-of-node twisting of the Galactic warp with classical Cepheids and suggested that the warp originated from the torques from the massive inner Galactic disk.

While the origin of the Galactic warp understandably remains a complex puzzle, simply defining the geometry of the warp is a problem that is also far from resolved, with a variety of potential models posited for its shape \citep{Romero-Gomez2019}. Even something as seemingly straightforward as the radius of the onset of the Galactic warp is still under debate. For example, \cite{Drimmel_Spergel2001} found the onset of the warp to lie $\sim1\ \text{kpc}$ {\it inside} the solar circle using a three-dimensional model for the Milky Way fitted to the far-infrared (FIR) and near-infrared (NIR) data from the COBE/DIRBE instrument, a result supported by \cite{Huang2018} using stars from TGAS-LAMOST.
\cite{Sch_Dehnen2018}, using \emph{Tycho}-\emph{Gaia} Astrometric Solutions (TGAS) data set, also claimed that the warp begins inside the solar circle. On the other hand, population synthesis models from \cite{Derriere2001} and \cite{Reyle2009} placed the onset of the Galactic warp at or outside the solar circle (see also \citealt{Romero-Gomez2019}, discussed further below).

In addition, the precession rate of Galactic warp is also unsettled.  \citet{Drimmel2000} claimed that the warp is precessing rapidly (about $25\ \text{km}\ \text{s}^{-1}\ \text{kpc}^{-1}$) in the direction of Galactic rotation, though the authors also acknowledge that the biased photometric distance caused the observed vertical motion to be smaller than their true values, mimicking the signal of precession.  On the other hand, \citet{Bobylev2010} analyzed the three-dimensional kinematics of about 82,000 Tycho-2 stars belonging to the red giant clump (RGC), and claimed that no significant precession of the warp is detected in the solar neighborhood. Most recently, \citet{Poggio2020} applied the precessing warp model from \citet{Drimmel2000} to Gaia DR2 data with warp starting radius, height and shape ($R_w$, $h_w$ and $\alpha$) fixed to values in previous studies, and report that the warp is precessing at $10.46\ \text{km}\ \text{s}^{-1}\ \text{kpc}^{-1}$, i.e., roughly half the rate found by Drimmel et al. Yet still more complicated are the definition of warp parameter dependencies as a function of stellar ages, which may bear on the evolution of the warp or on the relative responses of different stellar populations to perturbations. For example, \citet{Amores2017}, using 2-Micron All Sky Survey \citep[2MASS;][]{Skrutskie2006} data and the Besançon Galaxy Model \citep{Czekaj2014}, identified a clear dependence of the thin disc scale length as well as the warp and flare shapes with age. 
Meanwhile, the recent availability of enormous samples of Milky Way stars with precise 3-D kinematics coming from the second data release of \gaia \citep[\gaia DR2;][]{GaiaDR2} has enabled much more comprehensive analyses of Galaxy dynamics over large ranges of Galactocentric radius, with the added means to estimate ages for field stars, and with much greater statistical robustness for both.  For instance, \citet{Poggio2018} using a combined sample of \gaia DR2 and 2MASS photometry found the presence of a warp signal in two stellar samples having different typical ages, and suggested that this means the warp is a gravitationally induced phenomenon.  Shortly thereafter, \citet{Romero-Gomez2019} used two populations of different ages --- young (OB-type) stars and intermediate-old age (red giant branch, RGB) stars --- selected from \gaia DR2 and reported different onset radii for the Galactic warp for each, namely 12-13 kpc for the young sample versus 10-11 kpc for the older sample.  These authors also report that the older sample reveals a slightly lopsided warp, i.e., the warp is not symmetric in shape about the plane, with a possibly twisted line of nodes.

One of the significant outcomes of this new capability in Galactic astronomy is the mapping of stellar motions --- and asymmetries in those motions --- across the Milky Way disk \citep[e.g.,][]{Kawata2018,Poggio2018,Corredoira2020}. Such kinematical asymmetries would be expected in the presence of a warp, but they can also explain smaller-scale features.  For example, \cite{Bennett2019} and \cite{Carrillo2019} each reported a combination of bending and breathing modes using stellar kinematics derived from \gaia astrometry, and confirmed that the Galactic disk is undergoing a wave-like oscillation with a dynamically perturbed local vertical structure within the solar neighborhood.  

Such oscillatory motions may also explain various low latitude substructures that reside in the {\it outer} Galactic disk, like the Monoceros ring \citep{Newberg2002}, Triangulum-Andromeda (TriAnd) \citep{Rocha-Pinto2004, Majewski2004}, A13 \citep{Sharma2010,Li2017} and other ring-like overdensities \citep{Penarrubia2005}, whose origins have long been debated. For example, Monoceros and TriAnd were originally thought to be low-latitude tidal debris from dwarf galaxies \citep{Chou2010,Sollima2011,Sheffield2014}. However, there is now mounting chemical and kinematical evidence that some of these overdensities belong to the disk of the Milky Way \citep[][Sales Silva et al., in prep.]{Bergemann2018,Hayes2018,SalesSilva2019} and represent concentrations of stars at the crests or troughs of ripple-like density waves in the Galactic disk or vertical oscillations of the Milky Way midplane at large Galactocentric radii that are excited by orbiting dwarf galaxies (e.g., \citealt{Kazantzidis2008,Newberg2017,Laporte2018}).  If these overdensities are related with the local vertical structure of the Milky Way disk, they may therefore provide further constraining power on the source of these perturbations.

In this study we use \gaia DR2 and APOGEE together with the StarHorse distance solutions \citep{Anders2019} to explore vertical and radial velocity patterns and structures in the kinematics of the Galactic disk and to use these features to characterize the onset radius and precession rate of the warp. In Section \ref{sec:data}, we describe the sources of our data, the distances adopted and conversion to the Galactocentric reference frame. In Section \ref{sec:kinematics}, we present several detected kinematical signatures in vertical and radial velocity, and in Section \ref{sec:model} we apply a simple model, based on the Jeans Equation, to characterize these findings. In Section \ref{sec:age}, we compare the responses to the galactic warp in four different age populations. In Section \ref{sec:conclusion}, we present the main conclusions from our analysis and outline prospects for building on the present work.

\section{Data} \label{sec:data}
The data in this paper come primarily from \gaia DR2 \citep{Gaia2016,GaiaDR2} and the Apache Point Observatory Galactic Evolution Experiment (APOGEE \& APOGEE-2, \citealt{Majewski2017}), part of SDSS-III \citep{Eisenstein2011} and SDSS-IV \citep{Blanton2017}. We use these two primary sources to generate two different data-sets for our analysis of the Galactic warp.

The first data-set uses information for the 7,224,631 stars down to $G \simeq$ 13 for which \gaia DR2 provides full \emph{6-dimensional} phase space coordinates: positions ($\alpha$, $\delta$), parallaxes ($\varpi$), proper motions ($\mu^{*}_{\alpha}=\mu_{\alpha}\cos\delta$, $\mu_{\delta}$), and radial line-of-sight velocities ($v_{\rm los}$)  \citep{Gaia_RV2018}. From that catalog, stars with suspect photometry and stars where the $v_{\rm los}$ measurement is based on fewer than four \gaia transits are removed. In addition, we decontaminate our sample of stars from the Large Magellanic Cloud (LMC) and Small Magellanic Cloud (SMC) by removing any sources within 5 degrees of the center of these systems.

\begin{figure}
    \centering
    \includegraphics[width=\columnwidth]{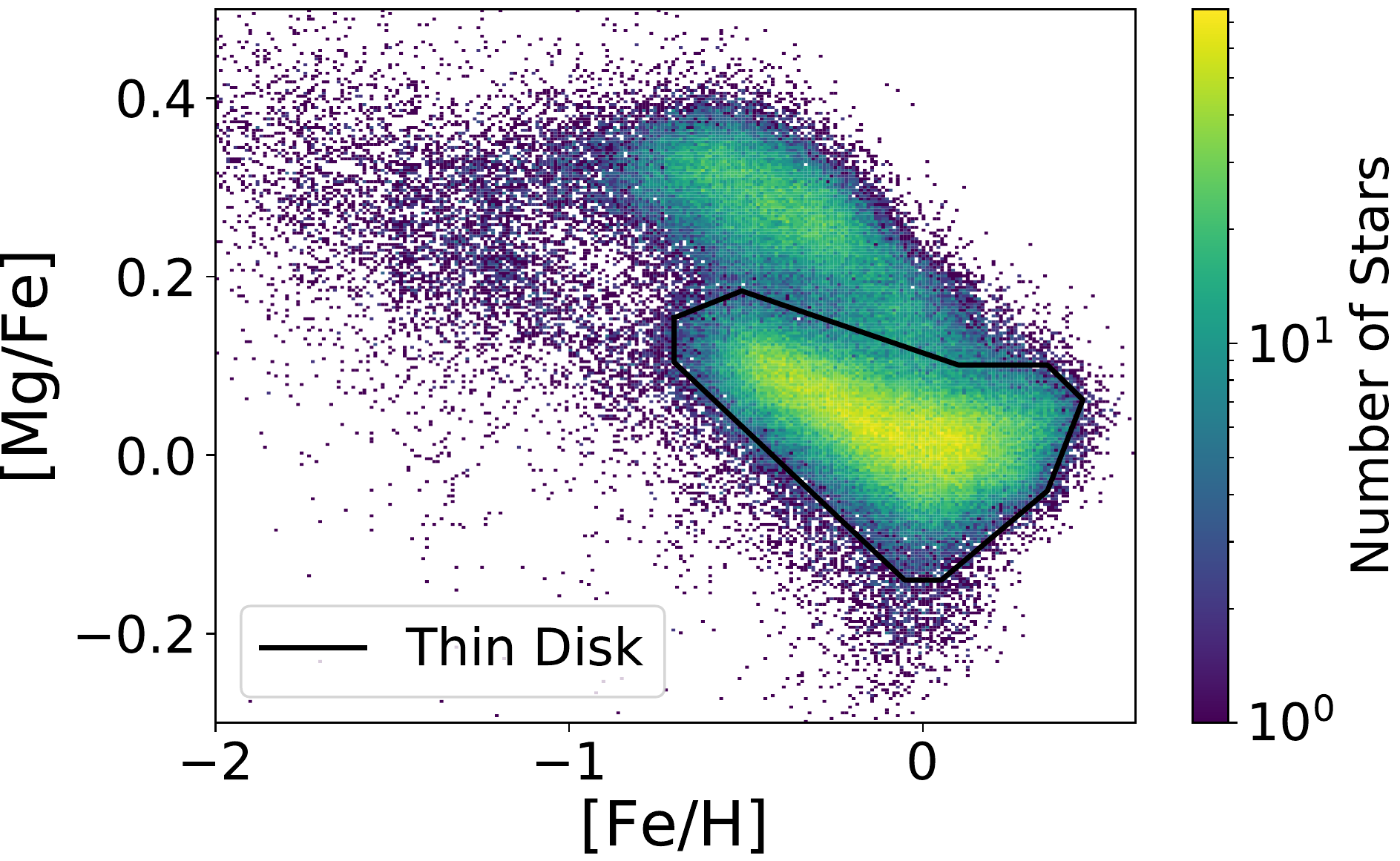}
    \caption{The [Mg/Fe]-[Fe/H] chemical plane from APOGEE DR16, from which we define our thin disk selection for our analysis. The colorbar represents the number of stars in each chemical bin (with yellow representing the highest density) and is on a logarithmic scale. Our thin disk selection is defined very conservatively by the solid line.}\label{fig:apogee_selection}
\end{figure}

Our second data-set is smaller, combining the proper motion, parallax and photometric information from \gaia DR2 with the chemical and radial velocity information from the latest public release of data from the APOGEE-2 survey, as contained in Sloan Digital Sky Survey (SDSS) Data Release 16 \citep[DR16][]{SDSS_DR16}. DR16 contains high resolution spectroscopic observations from both the Northern and Southern Hemispheres taken with the twin APOGEE instruments \citep{Wilson2019} on the Sloan 2.5-m \citep{Gunn2006} and the du Pont 2.5-m  \citep{Bowen1973} telescopes, respectively.  Individual stellar atmospheric parameters and chemical abundances are derived from the APOGEE Stellar Parameter and Chemical Abundance Pipeline (ASPCAP, \citealt{Garcia_Perez2016}). For SDSS DR16, ASPCAP has been updated to use a grid of MARCS stellar atmospheres (J\"onsson al. 2020) and a new $H$-band line list from Smith et al. (in prep.), all of which are are used to generate a grid of synthetic spectra against which are compared the target spectra to find the best match \citep[e.g.,][]{Zamora2015}. From the full APOGEE sample we require all sources to have the APOGEE \verb STARFLAG  and \verb ASCAPFLAG  set to ``0'' and to have an effective temperature between 3700K and 5500K. A further restriction in [Fe/H] and [Mg/Fe] was made to only keep stars having chemistry characteristic of stars in the thin disk \citep[see e.g.][]{Bensby2014, Hayes2018}, as illustrated and defined in \Cref{fig:apogee_selection}. The adopted thin disk selection criterion is defined very conservatively, to limit contamination by non-thin-disk stars.

\begin{figure*}
    \centering
    \subfloat[\gaia X-Y]{
        \includegraphics[width=0.49\textwidth]{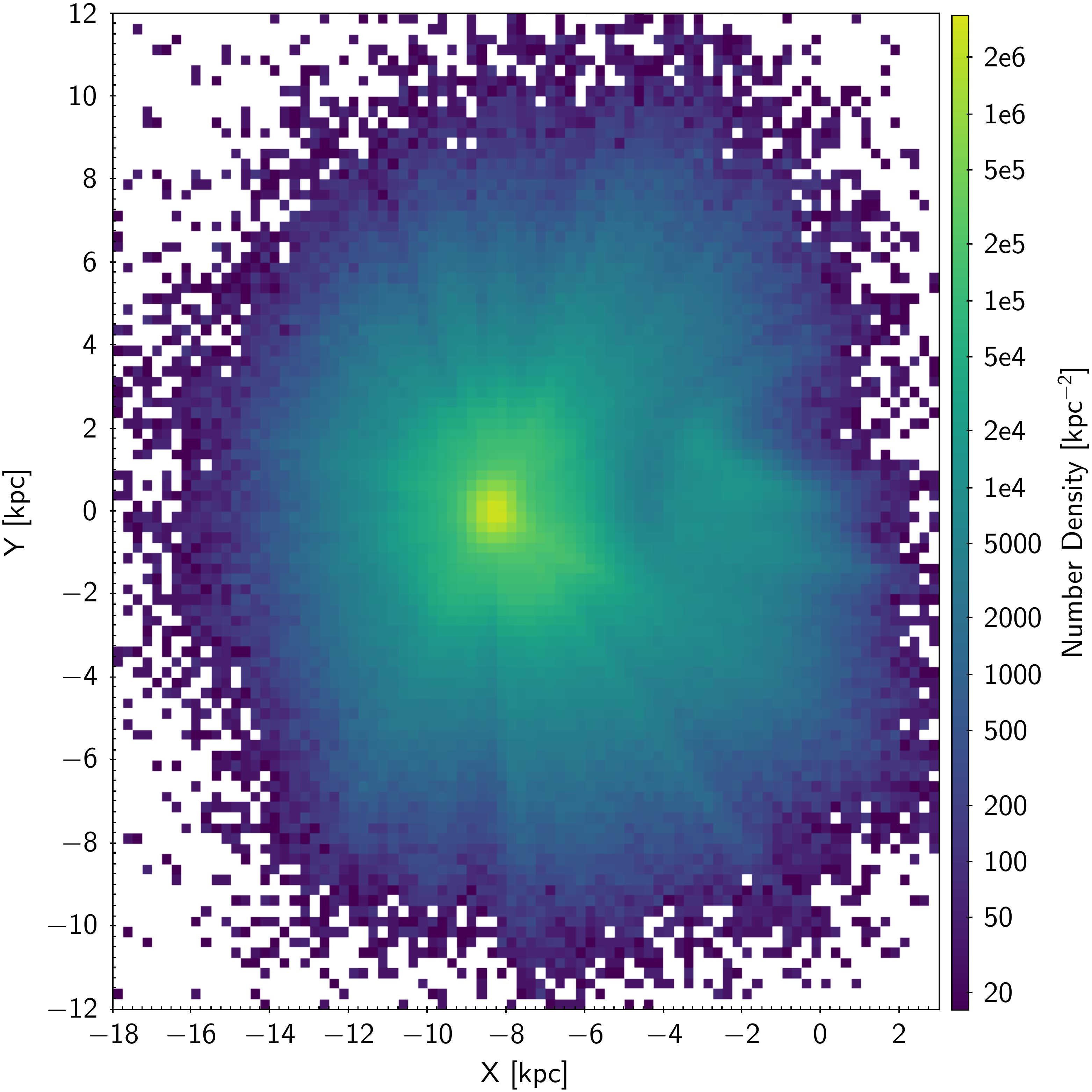}
        \label{fig:gaia_xy}
    }
    \hspace*{\fill}
    \subfloat[APOGEE X-Y]{
        \includegraphics[width=0.49\textwidth]{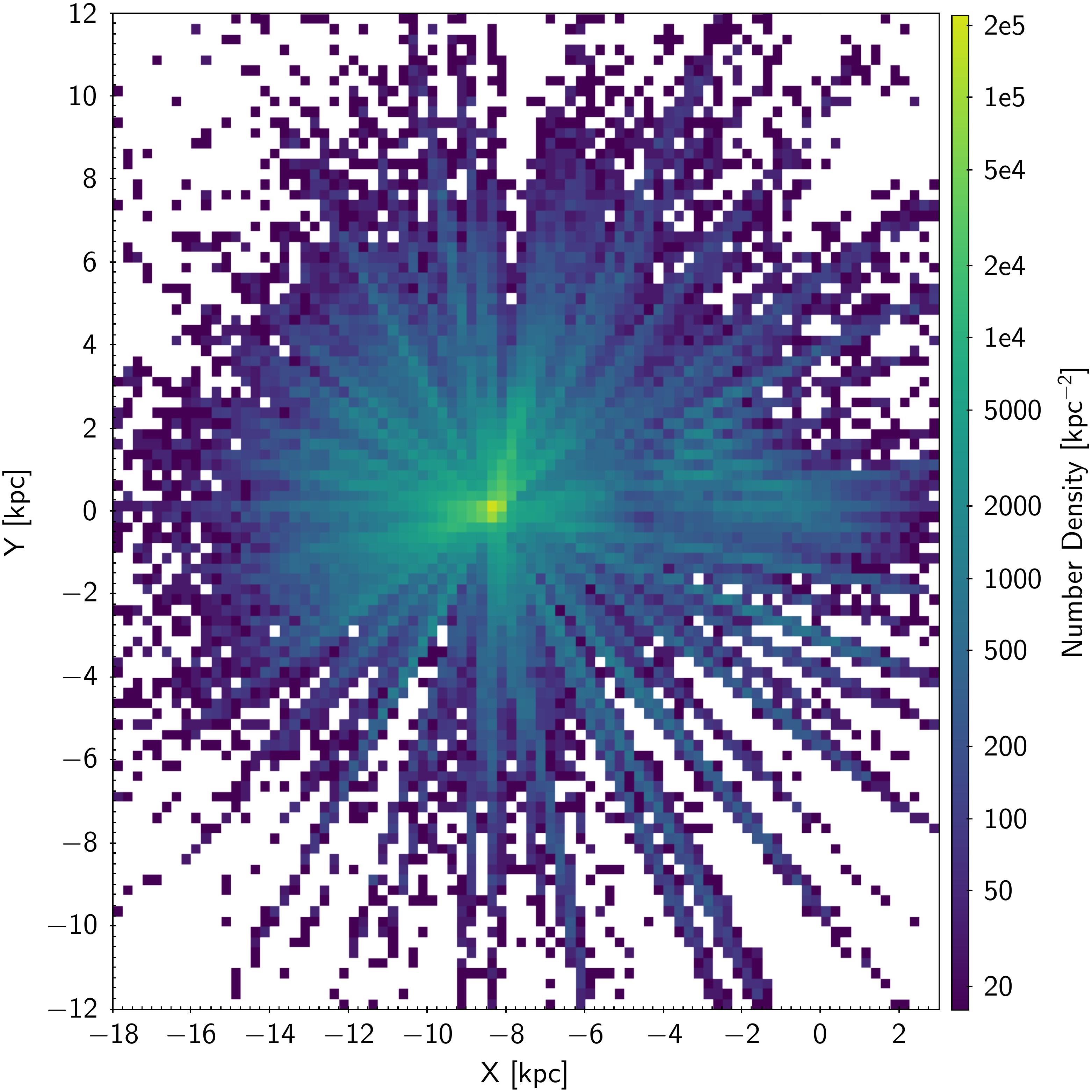}
        \label{fig:apogee_xy}
    }\\[-2ex]
    \subfloat[\gaia X-Z]{
        \includegraphics[width=0.49\textwidth]{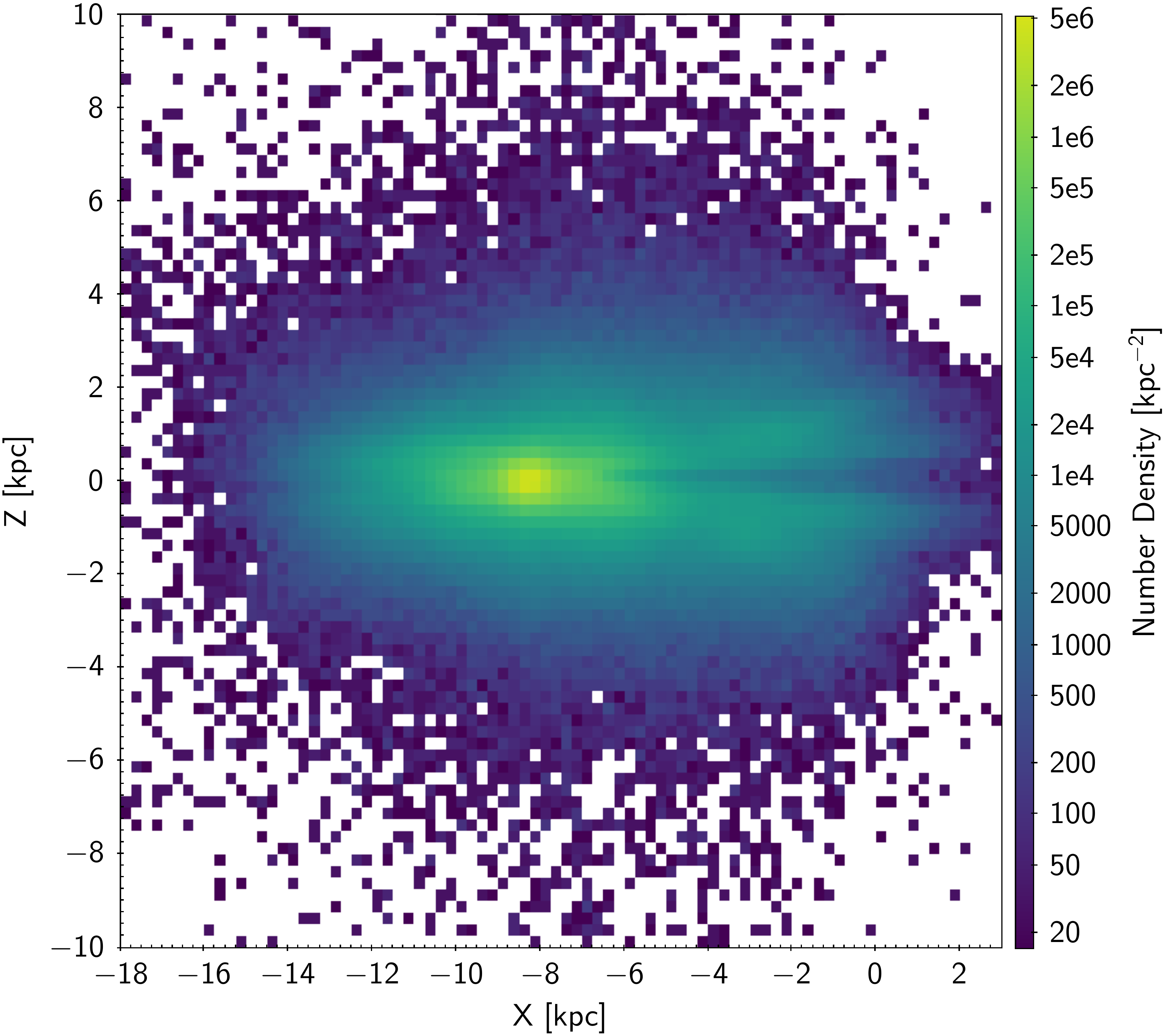}
        \label{fig:gaia_xz}
    }
    \hspace*{\fill}
    \subfloat[APOGEE X-Y]{
        \includegraphics[width=0.49\textwidth]{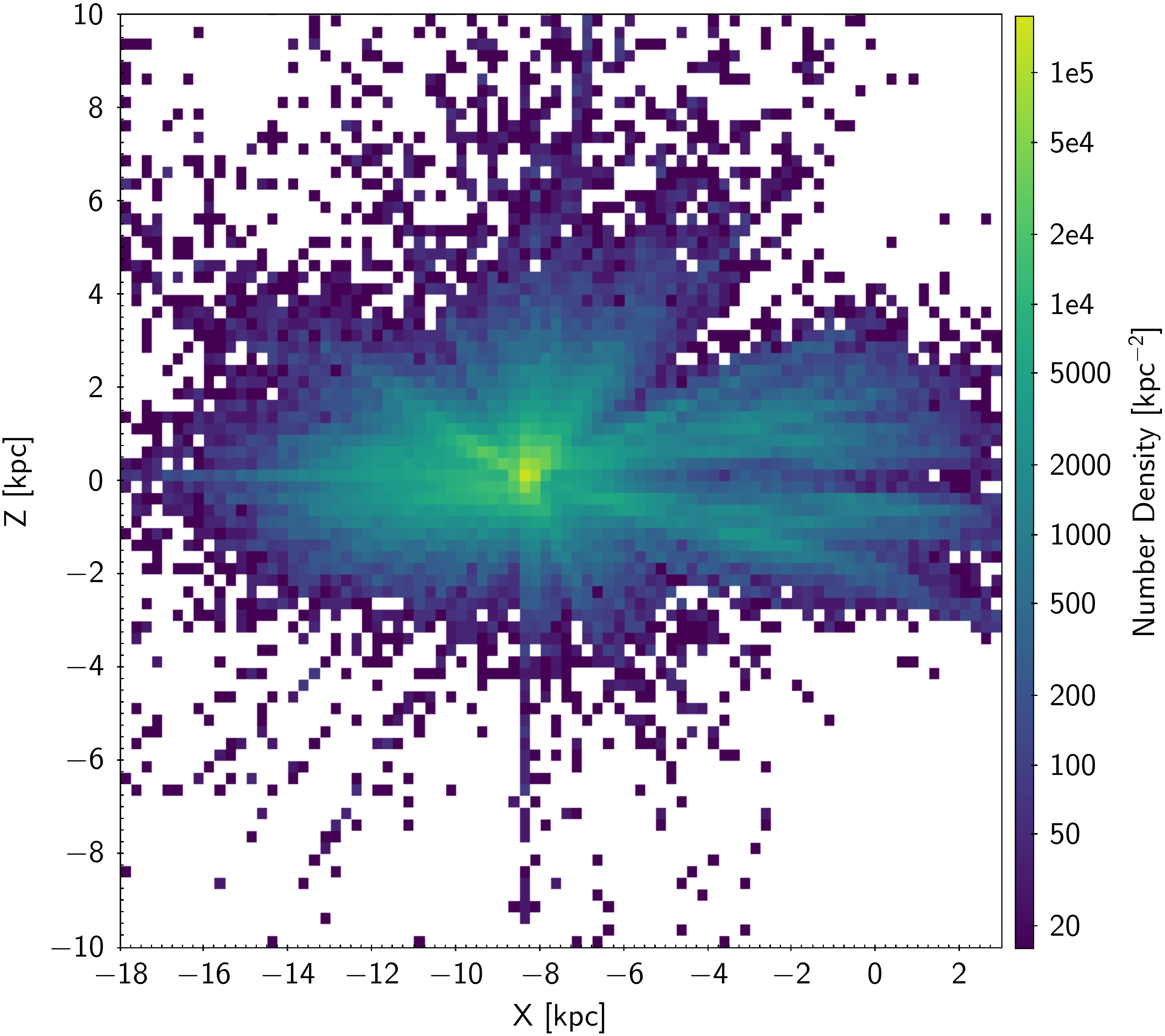}
        \label{fig:apogee_xz}
    }
    \caption{The spatial distributions of the {\it Gaia} (left) and {\it Gaia}-APOGEE (right) data-sets used in this paper, in Galactocentric coordinates. Each pixel represents $0.25\times0.25\ \text{kpc}^2$.  The top panel shows the $X_{GC}$-$Y_{GC}$ projection onto the Galactic plane, while the lower panels show projections onto the $X_{GC}$-$Z_{GC}$ plane. Similar plots are seen in \citet{Anders2019} and \citet{Queiroz2020}.}\label{fig:dist}
\end{figure*}

In this study we use distances derived through Bayesian inference using the StarHorse code \citep{Queiroz2018}.  StarHorse combines precise parallaxes and optical photometry delivered by \gaia DR2 with the photometric catalogues of Pan-STARRS1 \citep{Chambers2016}, 2MASS \citep{Skrutskie2006}, and AllWISE \citep{Wright2010}, aided by the use of informative Galactic priors \citep{Santiago2016,Queiroz2018}. For the APOGEE data-set we use the StarHorse distances and extinctions from  the APOGEE-2 DR16 StarHorse Value Added Catalog \citep{Queiroz2020}. The latter combines high-resolution spectroscopic data from APOGEE DR16 with the broad-band photometric data from the above sources and the \gaia DR2 parallaxes. Following the recommendation in \citet{Queiroz2020}, we adopt the combination of \verb SH_GAIAFLAG=="000"  and \verb SH_OUTFLAG=="00000"  to filter out stars that have a problematic \gaia photometric or astrometric solution or a troublesome StarHorse data reduction.

The mean distance uncertainties for stars in our \gaia/{\it Gaia}-APOGEE samples is $0.24/0.42\ \text{kpc}$, and the mean relative uncertainties of distance are $\sim 8/ 10\%$. The mean uncertainty for the proper motions are $0.06/0.06\ \text{mas}\ \text{yr}^{-1}$ for each the right ascension (RA) and declination (Dec) directions.  Meanwhile, the mean uncertainties for the radial velocities are $1.69\ \text{km}\ \text{s}^{-1}$ for those coming from \gaia, and $0.21\ \text{km}\ \text{s}^{-1}$ for those from APOGEE. The total number of stars in the {\it Gaia} and {\it Gaia}--APOGEE data-sets are 5,460,265 and 179,571, respectively.

The Galactocentric coordinate system adopted in this paper is right-handed, with the Sun at $(X, Y, Z) = (-8.12, 0, 0.02)$ kpc \citep{GravCollab2018, Bennett2019}, a Local Standard of Rest (LSR) velocity $V_{LSR}=233.4\ \text{km}\ \text{s}^{-1}$ \citep{Reid2004, GravCollab2018}, and a solar velocity relative to the LSR of $V_{\odot}=(12.9, 12.24, 7.78)\ \text{km}\ \text{s}^{-1}$ \citep{Drimmel2018}. Note that in this adopted Galactocentric Cartesian coordinate system, Galactic rotation converts to a negative azimuthal velocity ($v_{\phi}$) when expressed in cylindrical coordinates. For this reason, in many of the figures presented below, we adopt $-L_z$ for the abscissa.

Under this coordinate system, the spatial distribution of stars in our samples is shown in \Cref{fig:dist}.  As may be seen, both of our samples have kinematical information extending to $\sim$10 kpc from the Sun, although most of the stars are concentrated in the disk, within $Z=\pm 3\ \text{kpc}$ of the Galactic plane. The presence of the Galactic bar and bulge starts becoming evident at $X \lesssim -4$ kpc. The presence of the Galactic bar and bulge starts becoming evident at $X<-4$ kpc in our data, as already reported by \citet{Anders2019}, \citet{Queiroz2020} and Queiroz et al. in prep using the StarHorse distance solution. As expected, the all-sky {\it Gaia} sample is more smoothly and completely distributed, while the {\it Gaia}-APOGEE sample shows the pencil-beam spikes corresponding to the field-by-field coverage of the APOGEE and APOGEE-2 surveys, as well as the more limited coverage in the Southern Hemisphere, where APOGEE only began observing more recently in APOGEE-2.

\section{Kinematical Structures and Patterns}\label{sec:kinematics}
\subsection{The General Trend and Ripples in Vertical Velocity} \label{sec:v_z}
The warp and its kinematical signature are expected to be more prominent towards the Galactic anticenter and evident by large-scale systemic stellar motions perpendicular 
to the plane \citep[e.g.,][]{Binney1992,Drimmel2000}. Our {\it Gaia} and {\it Gaia}--APOGEE samples, in combination with the StarHorse distances, allow us to characterize the stellar vertical motion over a large range of Galactocentric radius, where we are able to explore to $R_{GC}\sim18\ \text{kpc}$. Here we study the kinematical signature of the Galactic warp in our two stellar samples, specifically by exploring the vertical velocity $v_{z}$ in the disk as a function of angular momentum ($L_z$) and Galactocentric radius ($R$), as was done previously by \citet{Sch_Dehnen2018,Huang2018}.  In addition, we look for any azimuthal asymmetries in these trends.

\begin{figure}
    \centering
    \includegraphics[width=\columnwidth]{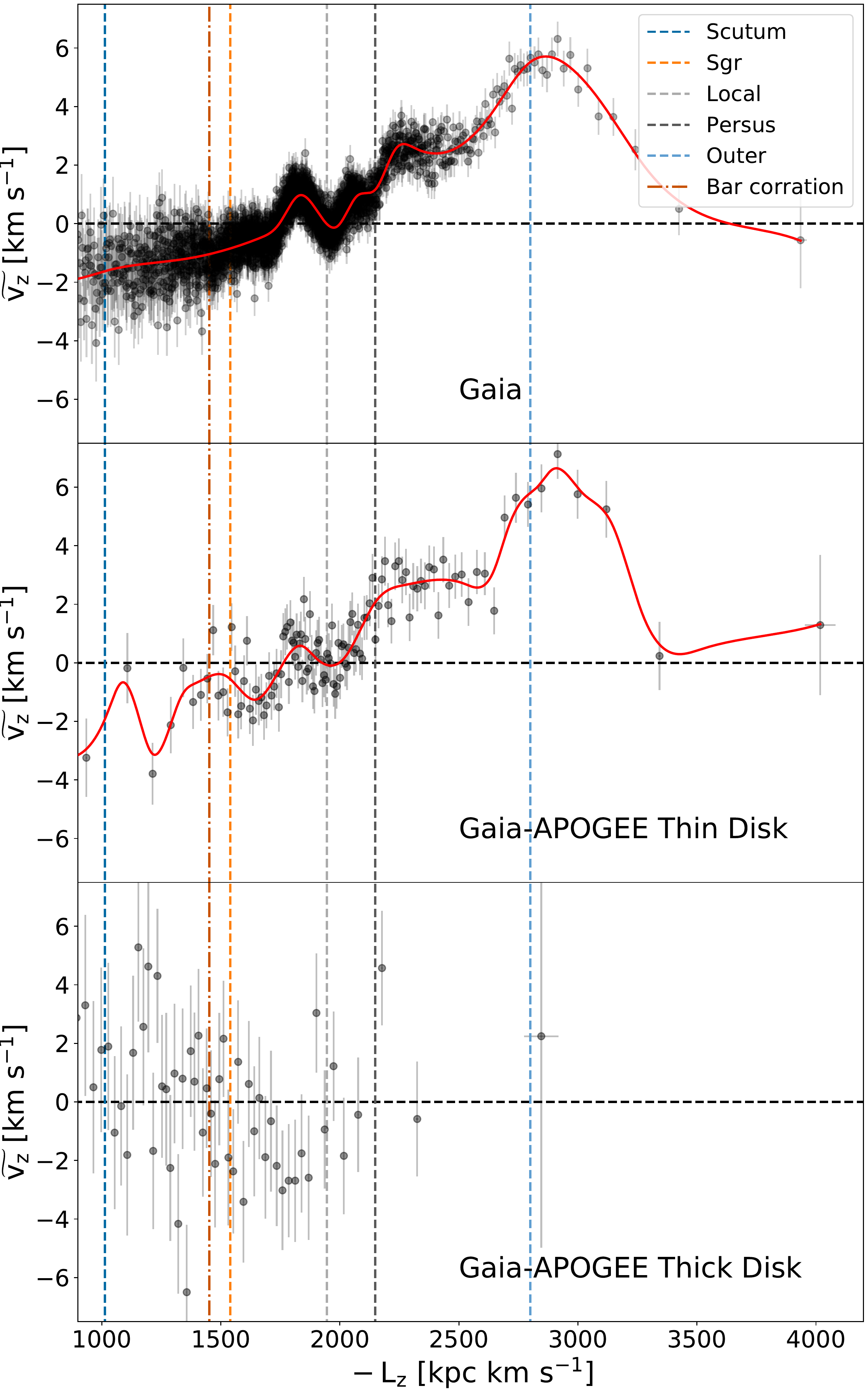}
    \caption{The vertical velocity ($\widetilde{V_z}$) versus angular momentum ($L_z$) for the {\it Gaia} ({\it top}) and {\it Gaia}-APOGEE thin disk ({\it middle}) and thick disk ({\it bottom}) samples.
    Stars are sequentially grouped into 2000-star bins for the {\it Gaia} sample, 1000-star bins for the {\it Gaia}-APOGEE thin disk sample, and 500-star bins for the {\it Gaia}-APOGEE thick disk sample (however, so that they would not be hindered by small sample size, the bins at largest $L_z$ contain 2265 stars for the {\it Gaia} dataset, 1571 stars for the {\it Gaia}-APOGEE thin disk dataset and 526 stars for the thick disk dataset). Each data point represents the median angular momentum and median vertical velocity ($\widetilde{V_z}$) for stars in that bin. Error bars represent the uncertainty of the median values, estimated through bootstrapping (see text). The solid red lines are smoothed trends to help visualize the data better. The dashed vertical lines indicate the angular momentum of spiral arms --- see discussion of the potential origin of the observed $\widetilde{V_z}$ ripples in Section 3.1.}\label{fig:vz_vs_lz}
\end{figure}
\begin{figure}
    \centering
    \includegraphics[width=\columnwidth]{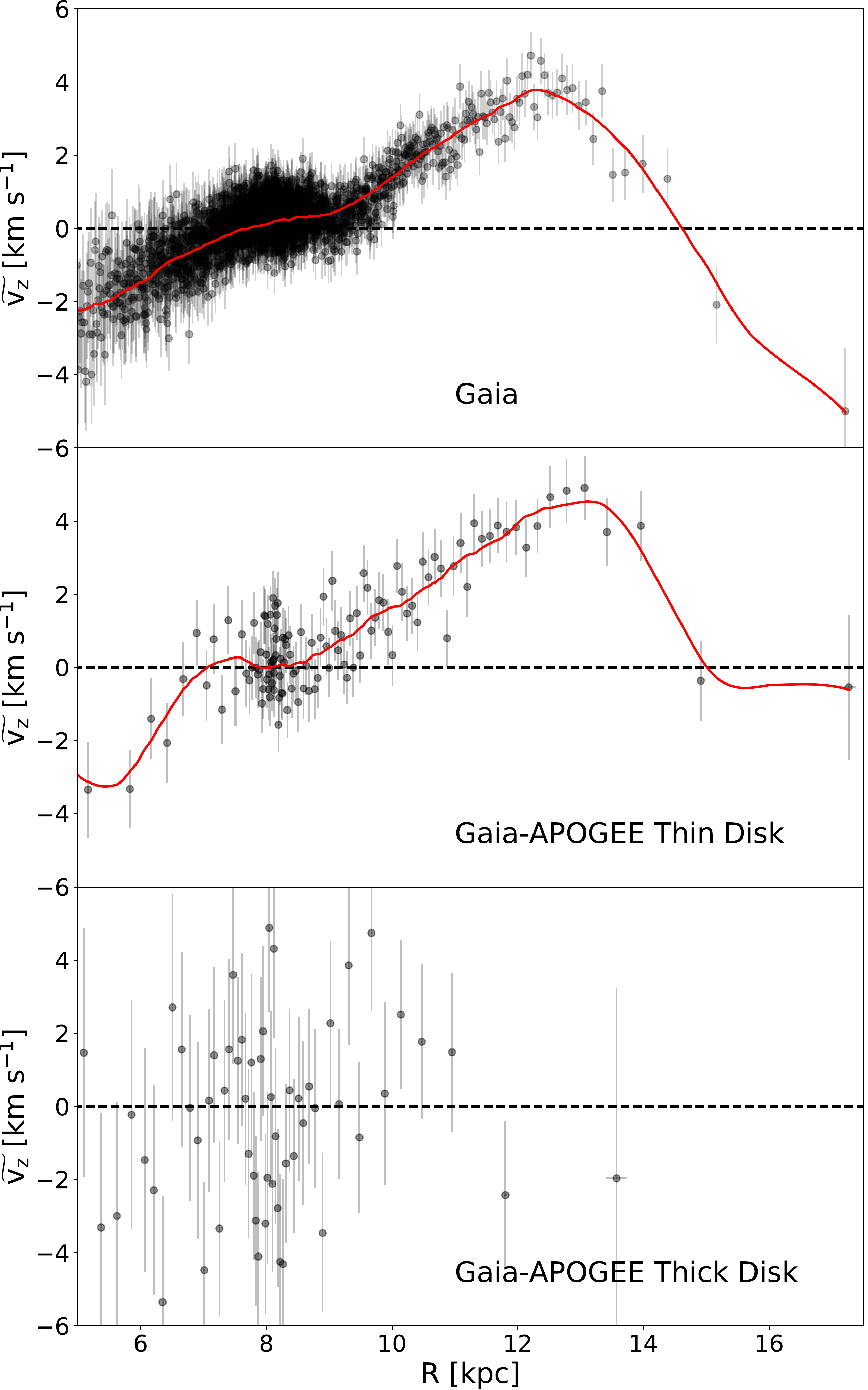}
    \caption{The same as in \Cref{fig:vz_vs_lz}, but now showing the
    vertical velocity ($v_z$) shown as a function of Galactocentric radius ($R$), with stars sorted and binned sequentially in $R$. As in Figure \ref{fig:vz_vs_lz}, the trend in the {\it Gaia} sample ({\it top panel}) is well-matched by the {\it Gaia}-APOGEE thin disk sample ({\it middle panel}), but not the thick disk sample ({\it bottom panel}).}\label{fig:vz_vs_r}
\end{figure}

\Cref{fig:vz_vs_lz} shows the run of $\widetilde{v_z}$ with $L_z$, for the {\it Gaia} data-set (in the left panel), and the chemically selected thin disk stars from the {\it Gaia}--APOGEE sample (in the right panel).  Stars in \Cref{fig:vz_vs_lz} are sorted and binned by angular momentum, with each point representing 2000 stars for the former data-set, and because the parent sample is smaller,
each point representing 1000 stars for the {\it Gaia}--APOGEE sample. The error bars represent the uncertainty of the median value, which have been estimated through bootstrapping: 1000 subsamples containing 80\% of the stars in each bin were randomly drawn and the standard deviation of the median of these subsamples were taken as error of the median.

The trend of the {\it Gaia} sample (top panel of \Cref{fig:vz_vs_lz}) strongly resembles that of the {\it Gaia}-APOGEE sample (middle panel).  Because the latter was deliberately chosen via chemistry (\Cref{fig:apogee_selection})  to select thin disk stars, we can conclude that the features shown in the larger {\it Gaia} sample are driven by thin disk stars. This conclusion is reinforced by the trend of the thick disk stars (bottom panel), which doesn't at all resemble the trend of the {\it Gaia} stars.

\Cref{fig:vz_vs_lz} shows that over a large range of $L_z$ the overall mean vertical velocity increases with $L_z$, starting at $-2\ \text{km}\ \text{s}^{-1}$ and peaking at 
around $+6\ \text{km}\ \text{s}^{-1}$. This velocity increase is more pronounced for values larger than $L_z> 1800\ \text{kpc}\ \text{km}\ \text{s}^{-1}$ and continues until $L_z\sim 2800\ \text{kpc}\ \text{km}\ \text{s}^{-1}$, after which $\widetilde{v_z}$ sharply declines. A general increasing trend of $\widetilde{v_z}$ with $L_z$ was also noted by \citet{Sch_Dehnen2018} and \citet{Huang2018}.  However, while these previous studies reported that the correlation between $\widetilde{v_z}$ and $L_z$ can be approximated by a rising linear fit over their entire sample, our more extensive radial coverage of disk kinematics reveals that the increasing trend is limited to $L_z \lesssim 1800\ \text{kpc}\ \text{km}\ \text{s}^{-1}$, beyond which $\widetilde{v_z}$ actually declines. {\it We believe that this entire global trend in $\widetilde{v_z}$ is the signature of the Galactic warp}, and we further characterize it and model it as such in Section \ref{sec:model}.

\Cref{fig:vz_vs_lz} also reveals, superposed on top of this general trend, higher frequency, wave-like ripples in $\widetilde{v_z}$ as a function of $L_z$.  The source of these $\widetilde{v_z}$ ripples is more elusive (and will be explored after examination of similar trends in radial velocity, discussed in Section \ref{sec:v_r}).  However, these ripples are less prominent in the vertical velocity versus Galactocentric radius plot (\Cref{fig:vz_vs_r}): Although the general trend of vertical velocity first increasing and then decreasing as Galactocentric radius increases is still evident, the ripples, especially those at solar radius, are smeared out in this representation.
We argue that the reason the ripples are present in \Cref{fig:vz_vs_lz} but not \Cref{fig:vz_vs_r} is because angular momentum is conserved for stars but for a given present-day radius, you have a mix of stars at different phases in their orbits. Therefore, after stars have made a few revolutions around the Galactic center, any initial spatial patterns would smear out when binned in Galactocentric radius, even while $L_{z}$ is preserved.

\begin{figure}
    \centering
    \includegraphics[width=\columnwidth]{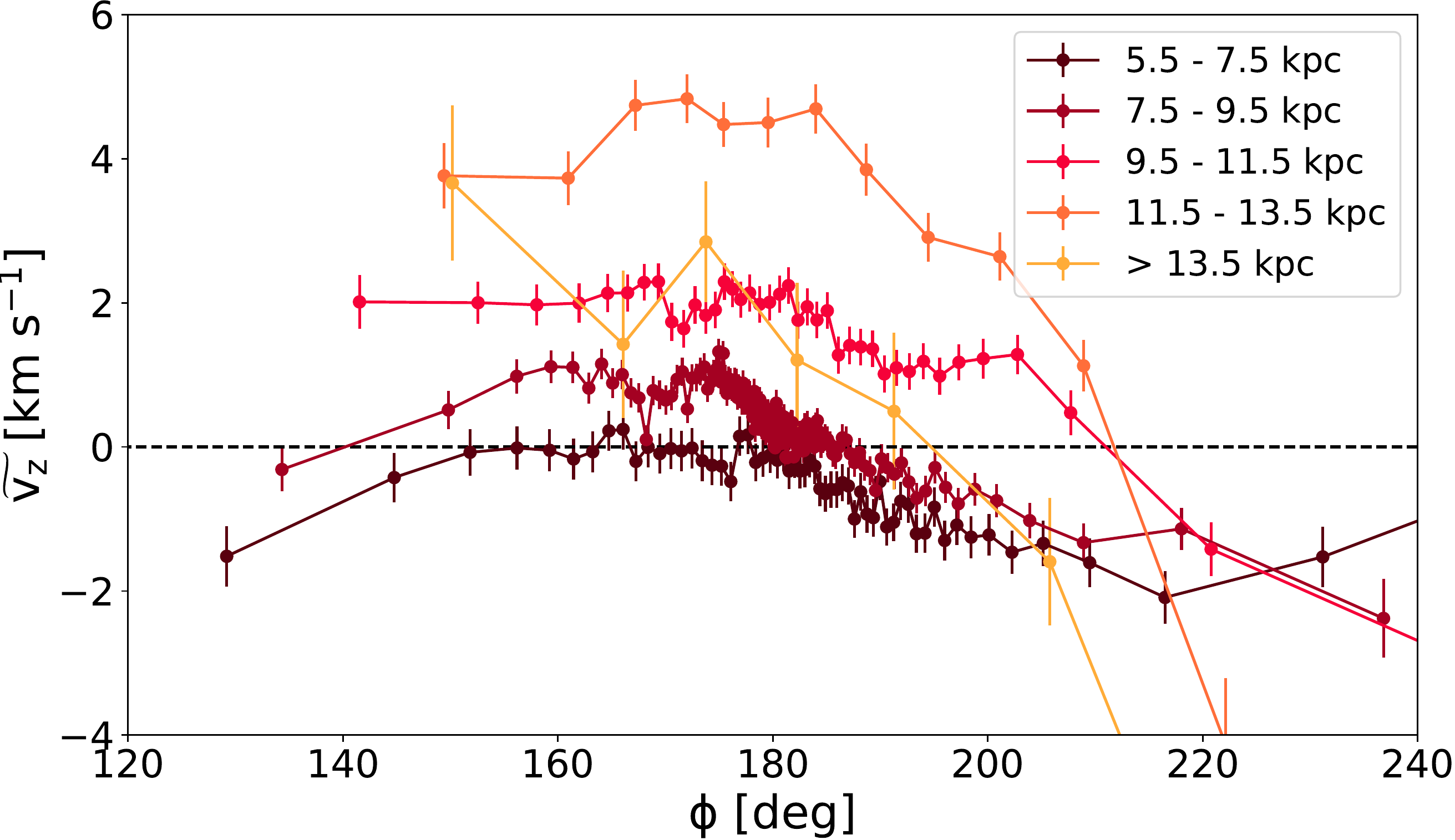}
    \caption{The vertical velocity versus Galactocentric azimuthal angle for the {\it Gaia} sample. The binning size is different for each radial bin: 20 000 stars/bin, 20 000 stars/bin, 10 000 stars/bin , 6000 stars/bin and 1500 stars/bin. The character of the trends for different radial annuli do not track one another; in particular, the rates of increase and decrease of $\widetilde{v_z}$ and the $\phi$ of maximum $\widetilde{v_z}$. These variations suggest that the warp may be lopsided.}\label{fig:phi_vz}
\end{figure}
\begin{figure}
    \centering
    \includegraphics[width=\columnwidth]{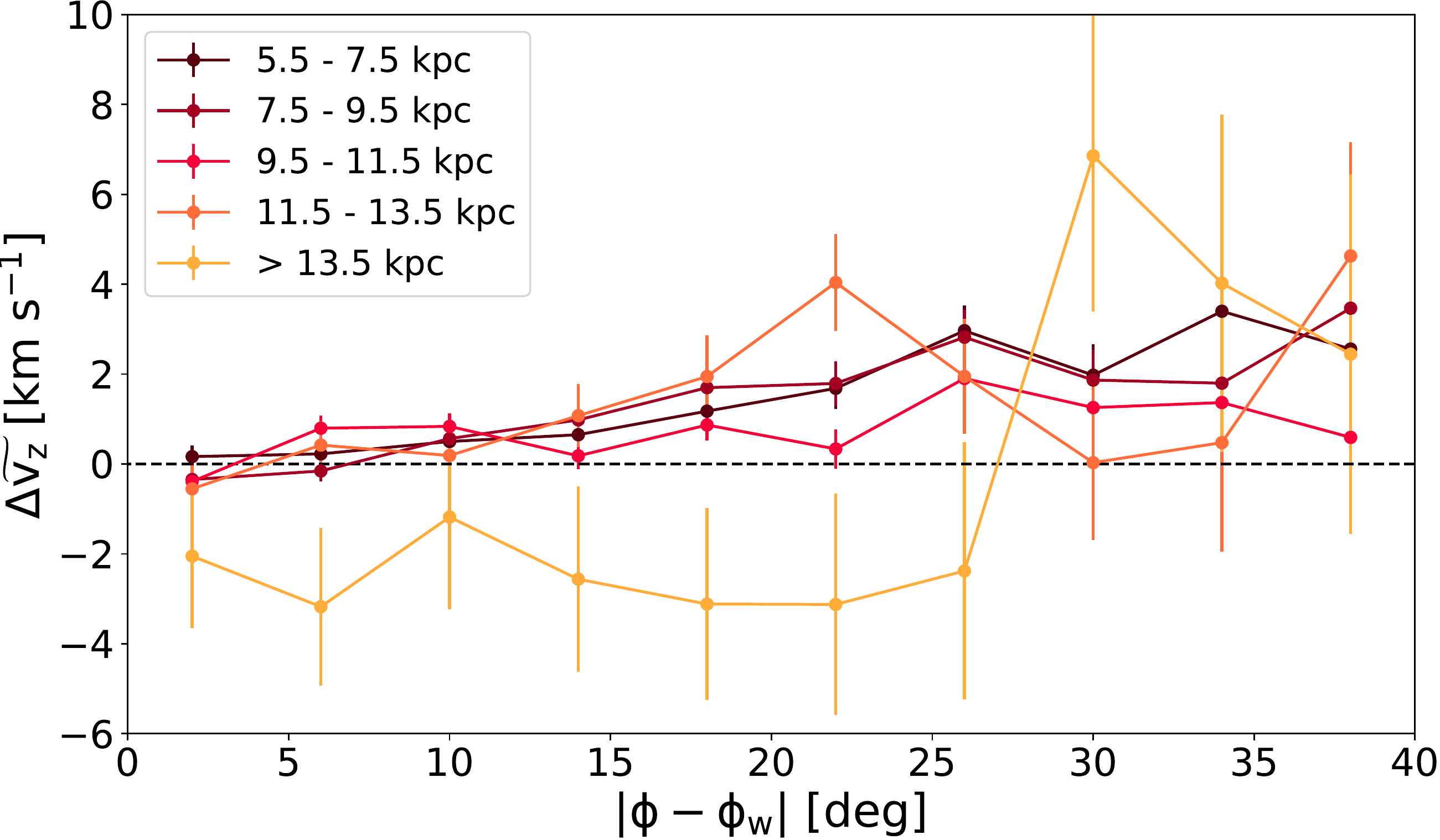}
    \caption{The vertical velocity differences between stars at $\phi < \phi_{peak}$ and $\phi>\phi_{peak}$ versus Galactocentric azimuthal angular separation from $\phi_{peak}$ for the {\it Gaia} sample. Stars are binned in 5$^{\circ}$ bins. $\phi_{peak}$ is determined through binning the stars with 10$^\circ$ bins and finding the bin with maximum vertical velocity.}\label{fig:phi_vz_absdel}
\end{figure}

An asymmetry in the Galactic H I warp has been extensively studied \citep{Burke1957,Baldwin1980,Henderson1982,Richter1994}, but whether there is a similar effect on the stellar disk is less understood. Recently, \citet{Romero-Gomez2019} reported asymmetry in the mean vertical distance of the stars from the Galactic plane about the warp line of nodes at $\phi \approx 180^{\circ}$, with the warp-down amplitude (at $\phi \gtrsim 180^{\circ}$) being larger than the warp-up amplitude (at $\phi \lesssim 180^{\circ}$), i.e., that the warp is lop-sided. Such differences in the amplitude of the spatial distribution may correlate to an asymmetry in the azimuthal variation of vertical velocity between the up and down sides of the warp. Furthermore, previous research also probed the possibility that the peak maximum vertical velocity is not in the anti-center direction \citep{Yusifov2004, Skowron2019, Li2020}.

To verify these presumptions, we plot the median vertical velocity as a function of Galactocentric azimuth angle, $\phi$, for different radial annuli in \Cref{fig:phi_vz}. At Galactocentric radii around the solar neighborhood, we see that the vertical velocity is relatively constant as a function of azimuthal angle; but at larger radii, we begin to see substantial differences in the vertical velocities at different azimuths.  In particular, the vertical velocity first increases as $\phi$ increases, reaches a maximum vertical velocity peak or plateau around $\phi\approx 170^{\circ}$, and then decreases with increasing $\phi$. Furthermore, the increasing and decreasing slopes of the vertical velocity with $\phi$ appear to be asymmetric about this peak or plateau, with a steeper decline in vertical velocity at $\phi > 170^{\circ}$ than the increase at $\phi < 170^{\circ}$. For a warp with an equal warp-up and warp-down amplitude, the vertical velocity should be symmetric about the longitude of peak vertical velocity; that this is not seen further indicates that the warp is lop-sided, as previous studies have identified using the altitude with respect to the Galactic plane at a given Galactocentric radius \citep[e.g.,][]{Marshall2006,Romero-Gomez2019}.

To illustrate further the kinematical lopsidedness of the Galactic warp, we directly measure the velocity asymmetry by subtracting the median vertical velocity of stars on one side of $\phi_{peak}$ from its complement on the other side at the same azimuthal separation for each radial annulus (\Cref{fig:phi_vz_absdel}), and $\phi_{peak}$ is estimated within each radial annulus by using a wider bin (10$^\circ$) between $160 < \phi< 200^\circ$.

\subsection{Ripples in Radial Velocity and their Potential Origin}\label{sec:v_r}
\begin{figure}
    \centering
    \includegraphics[width=\columnwidth]{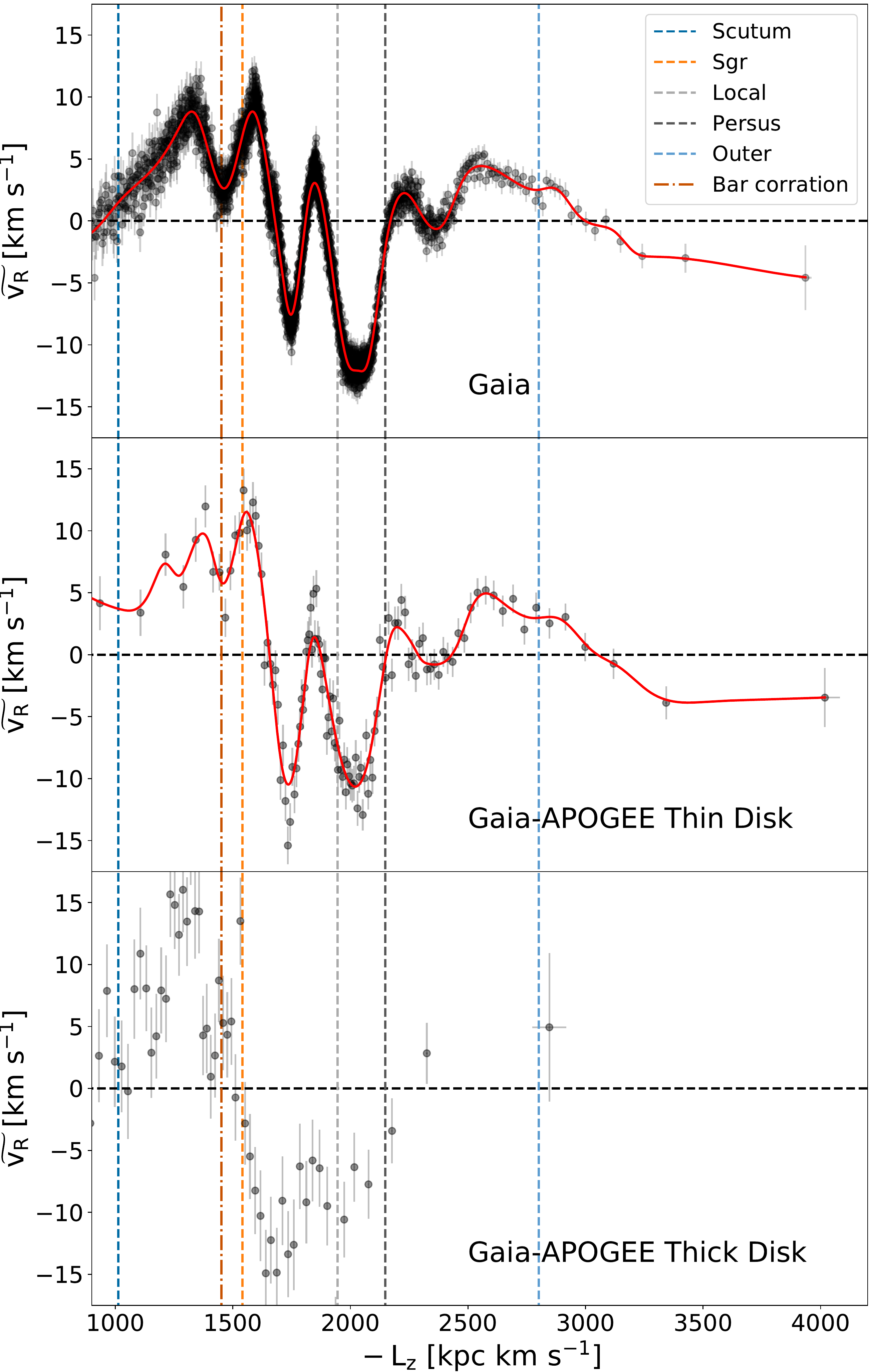}
    \caption{The same as in \Cref{fig:vz_vs_lz}, but now showing the radial velocity ($v_{R}$) versus angular momentum ($L_{z}$) for the {\it Gaia} ({\it top}), {\it Gaia}-APOGEE thin disk ({\it middle}), and {\it Gaia}-APOGEE thick disk ({\it bottom}) samples.}\label{fig:vr_vs_lz}
\end{figure}

\begin{figure}
    \centering
    \includegraphics[width=\columnwidth]{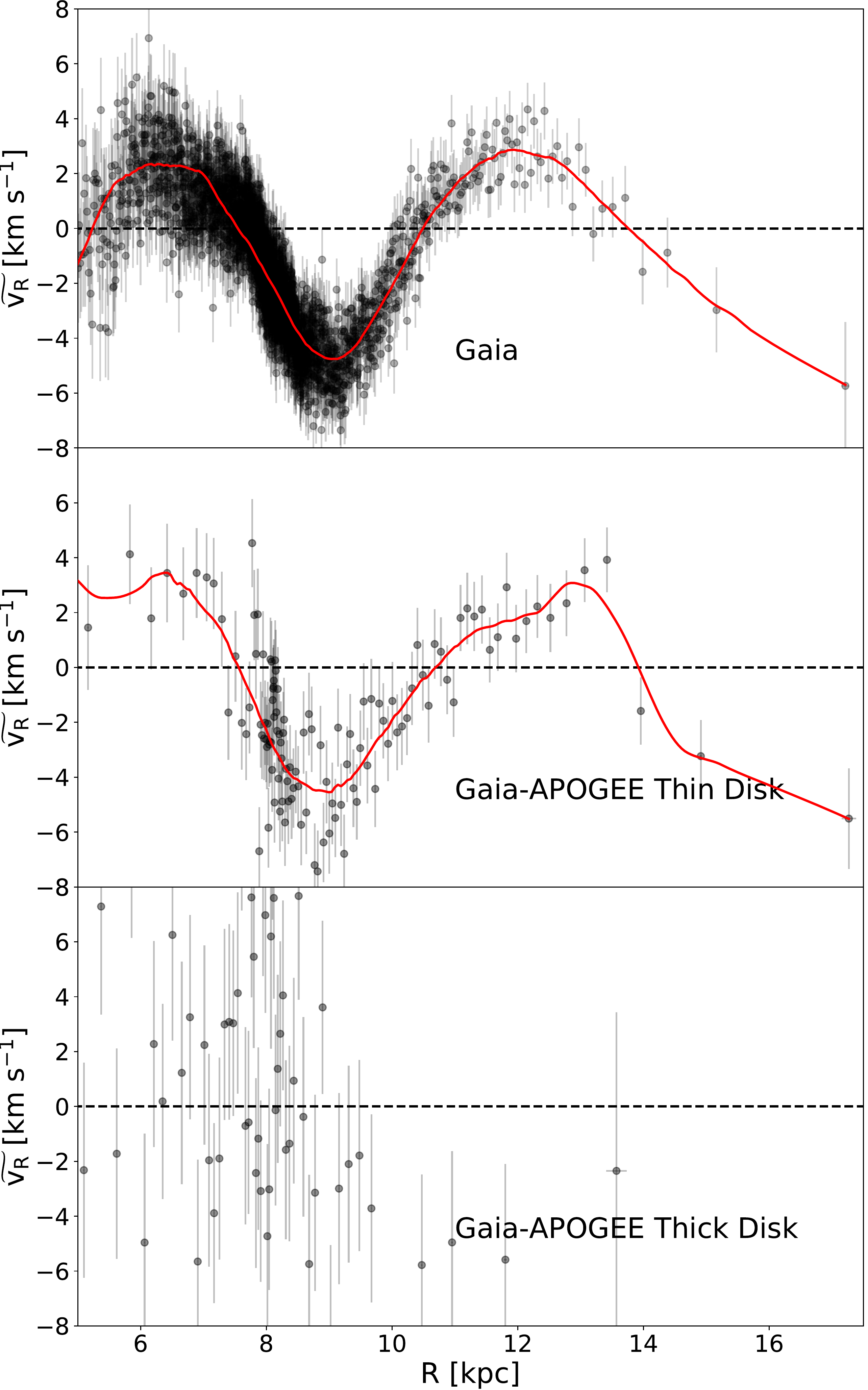}
    \caption{The same as in \Cref{fig:vz_vs_r}, but now showing the radial velocity ($v_{R}$) versus Galactocentric radius ($R$).}\label{fig:vr_vs_r}
\end{figure}

\Cref{fig:vr_vs_lz} and \Cref{fig:vr_vs_r} show the complementary stellar radial motions with respect to $L_{z}$ and $R$, respectively. Here, again, a pattern of ripples is seen, and they are even more dramatic, reaching more extreme velocity amplitudes. As was observed with the $v_{z}$ trends (i.e., \Cref{fig:vz_vs_lz} and \Cref{fig:vz_vs_r}) (a) the trends of the {\it Gaia} sample are best matched by the {\it Gaia}-APOGEE thin disk sample rather than the thick disk sample, and (b) the $v_{R}$ ripples seen when plotted as a function of $L_{z}$ are smeared out when plotted as a function of Galactocentric radius.  Such a kinematical pattern for $v_R$ was also reported for very young OB stars alone when viewed with respect to Galactocentric radius \citep{Cheng2019}.


Two possible mechanisms have previously been proposed to lead to such observed ripples.  One explanation for these localized features is that they may be related to spiral arm perturbations, where the mass enhancements associated with spiral density waves can gravitationally scatter stars \citep[e.g.,][]{Jenkins1990}. To illustrate the potential connection of these oscillations to spiral arms, in \Cref{fig:vz_vs_lz} we indicate the angular momentum values of the known Milky Way spiral arms, calculated as follows: First we take the parameters that characterize these spiral arms from \citet{Reid2014}. Then, a number of equally-spaced points were generated within the Galactocentric radius and azimuthal angle range of each spiral arm.  The standard rotation curve from \citet{Bovy2015} is assumed and used to calculate the azimuthal velocity. The angular momentum for each point is then computed and a variety of statistics (median, standard deviation, min/max values) generated to describe each spiral arm (see \Cref{tab:1}). We also show the $L_z$ position corresponding to the Galactic bar, where we assume a pattern speed of $\Omega=39.0\ \text{km}\ \text{s}^{-1}\  \text{kpc}^{-1}$ \citep{Portail2017}.

After performing this simple exercise, we find (\Cref{fig:vz_vs_lz}) that even though the nominal Perseus and Outer arms correspond to local maximum of the vertical ripples, the Scutum, Sgr and Local arms do not. In terms of radial motion (\Cref{fig:vr_vs_lz}), there are many $v_{R}$ features that appear to be matched to corresponding $v_{z}$ features at the same $L_{z}$, and some correlations between the $L_{z}$ positions of some ripples and those characteristic of spiral arms can be seen --- in  particular, once again, between the $L_{z} \sim 1950$ kpc km s$^{-1}$ valley and the Local Spiral Arm and the peak at $L_{z} \sim 2150$ kpc km s$^{-1}$ with the Perseus Spiral Arm, but in this case no correlation between the Outer Spiral Arm and a $v_{R}$ feature is seen. We expect spiral arms to couple more tightly with the {\it radial} motions of stars, making the radial velocity dispersion significantly greater than the velocity dispersion perpendicular to the plane \citep[e.g.,][]{Jenkins1990,Aumer2016}. However, while most spiral arms correspond to maximum positive velocity, the local arm is at the point of maximum negative velocity. It is also apparent that some ripples visible in these figures {\it do not} correlate with known spiral arm patterns, while some spiral arms (in particular, those at smaller $L_z$) {\it do not} match observed ripples.  While these discrepancies might suggest that the ripples are not (or not entirely) generated by spiral arms, such lack of one-to-one correlation may also reflect shortcomings of the above illustrative exercise and the many uncertainties and simple assumptions used to generate it.

\begin{table*}
    \caption{Summary of the vertical angular momentum properties for different known Milky Way spiral arms, calculated as described in Section \ref{sec:v_z}. The second and third columns are the median angular momentum and corresponding standard deviation for each spiral arm. The fourth and fifth columns are the minimum and maximum angular momentum of the points associated with each spiral arm.}\label{tab:1}
    \centering
    \begin{tabular}{ccccc}
    \hline
    Spiral arm & $\widetilde{L_z}$ ($\text{kpc}\ \text{km}\ \text{s}^{-1}$) & $\sigma_{L_z}$ ($\text{kpc}\ \text{km}\ \text{s}^{-1}$) & minimum $L_{z}$ ($\text{kpc}\ \text{km}\ \text{s}^{-1}$) & maximum $L_{z}$ ($\text{kpc}\ \text{km}\ \text{s}^{-1}$) \\
    \hline\hline
    Scutum & 1012 & 202 & 661 & 1427 \\
    Sagittarius & 1540 & 85 & 1374 & 1705 \\    
    Local & 1945 & 98 & 1755 & 2136 \\  
    Perseus & 2148 & 191 & 1783 & 2541 \\  
    Outer & 2800 & 225 & 2364 & 3264 \\
    \hline
    \end{tabular}
\end{table*}

Another mechanism to produces the ripples that has been proposed is perturbations of the disk caused by satellite galaxies. It has been suggested that the Galactic disk oscillates vertically due to radially propagating waves --- i.e., bending waves caused by the passing of orbiting Milky Way satellites \citep{Hunter_Toomre1969}, such as the Sagittarius dSph \citep{Ibata1998, Laporte2018, Darling2019}. Some success in modeling these features in the stellar disk (but for more limited empirical mappings of Milky Way features than presented here) has been shown by \citet{Widrow2014} and \citet{Laporte2019}, who invoke a semi-analytical prescription and N-body simulation of the Sagittarius dwarf galaxy interacting with the Galactic disk to explain the oscillatory disk star patterns. In both cases, regardless of the mass of the impactor, changes in vertical velocity on the scale of $5\ \text{km}\ \text{s}^{-1}$ within $R<20\ \text{kpc}$ in the anti-center direction are observed, especially in \citet{Laporte2019}, where their model L2 exhibits a strikingly similar overall trend to that of the observations, with vertical velocity increasing with Galactocentric radius over $5<R<10\ \text{kpc}$ to a maximum value $\sim 5\ \text{km}\ \text{s}^{-1}$, and then decreasing with Galacocentric radius over radii $13<R<20\ \text{kpc}$. 

Meanwhile, in N-body simulations of passages of massive satellite galaxies around a Milky Way-like disk galaxy, \citet{Donghia2016} find an increasing vertical velocity over $5<R<10\ \text{kpc}$ and decreasing vertical velocity for $13<R<20\ \text{kpc}$, as well as some smaller ripples within $7<R<10\ \text{kpc}$. Ripples in the radial dimension as large as $20\ \text{km}\ \text{s}^{-1}$ have been detected in the \citet{Donghia2016} simulations that are attributable to the Galactic warp itself.

At present, we offer no definitive explanation for the fine structure seen in the kinematical trends in Figures \ref{fig:vz_vs_lz} and \ref{fig:vr_vs_lz}.  Like \citet{Sch_Dehnen2018}, \citet{Huang2018} and \citet{Friske2019}, we point out these high frequency kinematical features  but do not offer a physical model to explain them.  On the other hand, we find that either (or both) the spiral arm and satellite perturbation scenarios seem viable.
For the remainder of the analysis here, we focus on attempting to describe the more global trends visible in Figures \ref{fig:vz_vs_lz}-\ref{fig:vr_vs_r}) --- in particular, the large scale trends one might expect to be produced by a large disk warp.

\section{Modeling the Global Properties of the Observed Vertical Disk Motions}\label{sec:model}
We can treat stars as a collisionless fluid and apply the first Jeans equation to link the kinematics of Galactic stars to their number density through the Collisionless Boltzmann Equation (CBE hereafter) \citep{Jeans1915,Henon1982}. A simple analytical model for the Galactic warp can be derived using the CBE \citep[e.g., Equation 11 in][]{Drimmel2000}, after adopting several simplifying assumptions.  Here we follow the \citet{Drimmel2000} approach, but without making as many simplifications. For example, Drimmel et al. assume there are no net radial motions, i.e., $v_R=0$.  However, our datasets binned in vertical angular momentum, $L_{z}$, and Galactocentric radius, $R$ (see \Cref{fig:vr_vs_lz} and \Cref{fig:vr_vs_r}, respectively), show an even larger velocity range in the radial direction (from $-5$ to $7\ \text{km}\ \text{s}^{-1}$) than in the vertical direction (from $-4$ to $4\ \text{km}\ \text{s}^{-1}$) at $R>6$ kpc. Therefore, we build a similar model to that of \citet{Drimmel2000}, but one that accounts for a non-zero radial motion, $v_R$. While our model attempts to take another step in degree of sophistication, it is still very simple and does not capture all of the possible physics.  In particular, it does not include warp lopsidedness.

The essence of the model is to treat the Galactic warp as a perturbation in the Milky Way disk. For an unperturbed (non-warped) disk, one can assume perfect axisymmetry for simplicity, eliminating the dependence of the unperturbed parameters on Galactocentric azimuthal angle, $\phi$. Moreover, one can assume that the unperturbed number density is a separable function with respect to Galactocentric radius $R$ and distance from the Galactic plane $z$:

\begin{equation}
n(R, z)=f(R)g(z)
\end{equation}
In this circumstance, the addition of a Galactic warp perturbation would only have an effect on the vertical direction. Namely, stars that reside at a given position $(R, \phi, z)$ are deviated by $z_0(R,\phi,t)$. Therefore, the perturbed number density of stars can be written as
\begin{equation}
n'(R,\phi,z)=n(R,z-z_0)=f(R)g(z-z_0)
\end{equation}
Accounting for the warp, according to \citet{Drimmel2000}, one could write $z_0$ as
\begin{equation}
z_0=h(R)\sin(\phi-\phi_w+\omega_p t)
\end{equation}
where $h(R)$ is the deviation from the Galactic mid-plane at a given Galactocentric radius $R$, $\phi$ is the Galactocentric azimuthal angle, $\phi_w$ is the line of nodes at present day ($t = 0$) --- i.e., where there is no vertical displacement ($z_0=0$) --- and $\omega_p$ is the precession rate of the warp.

An analytical form of $h(R)$ is given in \citet{Drimmel2000}, but  \citet{Romero-Gomez2019} pointed out that a model with an ending radius of the Galactic warp and flexible exponents in $h(R)$ would reproduce observed kinematical patterns better. Thus, we adopt a new analytical form of $h(R)$ by merging the models from these two above-mentioned sources:
\begin{equation}
h(R)=\begin{cases}
      0 & R\leq R_1 \\
      \frac{R_w}{R_2-R_1}(R-R_1)^\alpha & R_1<R\leq R_2 \\
      h(R_2)+\frac{dh}{dR}|_{R=R_2}(R-R_2) & R>R_2 \\
   \end{cases}
\end{equation}
where $R_1$ is the starting radius of the warp, $R_2$ is the ending radius of the warp, $R_w$ is a scale factor for the warp height, and the exponent $\alpha$ characterizes the shape of the warp. We can write the first Jeans Equation in cylindrical coordinates as 
\begin{equation}
\frac{\partial n'}{\partial t}+\frac{\partial(n'\overline{v_R})}{\partial R}+\frac{1}{R}\frac{\partial(n'\overline{v_\phi})}{\partial \phi}+\frac{\partial(n'\overline{v_z})}{\partial z}=0.
\end{equation}
However, because $v_\phi$ is not perturbed by the warp, one can still apply the axisymmetric condition, so that $\frac{\partial\overline{v_\phi}}{\partial \phi}=0$. We also adopted the assumption $\frac{\partial\overline{v_z}}{\partial z}=0$ made by \citet{Drimmel2000} for simplicity. After using the product rule of derivatives and applying the above assumptions, one finds
\begin{eqnarray}
[\overline{v_z}-(\frac{\overline{v_\phi}}{R}+\omega_p)h(R)\cos(\phi-\phi_w+\omega_p t)-\nonumber\\
\overline{v_R}\frac{dh}{dR}\sin(\phi-\phi_w+\omega_p t)] f(R)\frac{dg}{dz}\nonumber\\
+\overline{v_R}\frac{df}{dR}g(z-h(R)\sin(\phi-\phi_w+\omega_p t))+\nonumber\\
\frac{\partial\overline{v_R}}{\partial R}n'=0
\end{eqnarray}
Unlike the more simplified treatment in \citet{Drimmel2000}, here the factors $n'$, $f(R)$ and $g(z)$ do not cancel out. We assume the initial mass function (IMF) is a constant across the entire galaxy, so that the number density of stars is directly proportional to the mass density of the stellar disk. 
From the similarity in behavior displayed between the {\it Gaia} versus {\it Gaia}-APOGEE samples in \Cref{fig:vz_vs_lz} we conclude that the outer disk, where the warp happens and which is the focus of our interest, is dominated by thin disk stars; thus we can safely assume the mass density follows a double exponential potential like that followed by thin disk stars. Thus, we assume
\begin{equation}
n(R, z)=n_0\exp(-\frac{|z|}{z_h}-\frac{R}{R_h})
\end{equation}
where $R_h$ is the scale length and $z_h$ is the scale height of the thin disk.  

Adopting this as the density and the assumption that $\phi_w=180^\circ$, one then obtains a final equation that links together the different components of velocity:
\begin{multline}
\frac{\partial\overline{v_R}}{\partial R}=\frac{\overline{v_R}}{R_h}+\frac{sign[z-z_0]}{z_h}\left[\overline{v_z}-\left(\frac{\overline{v_\phi}}{R}+\omega_p\right)h(R)\cos\theta\right.\\
\left.-\overline{v_R}\frac{dh}{dR}\sin\theta\right]
\end{multline}
where $\theta=\phi-\phi_w+\omega_p t$. Since we assume the distribution of the stellar population is symmetric about $z_0$, the final observed vertical velocity is the average of those with $z>z_0$ (where $sign[z-z_0]=1$) and $z<z_0$ (where $sign[z-z_0]=-1$), yielding
\begin{equation}
    \overline{v_z}=\left(\frac{\overline{v_\phi}}{R}+\omega_p\right)h(R)\cos\theta+\overline{v_R}\frac{dh}{dR}\sin\theta
\end{equation}

The free parameters in our model arethe Galactocentric radius where the warp starts and ends ($R_1$ and $R_2$, respectively), the scale height of the warp ($R_h$), and the precession speed of the warp ($\omega_p$). This is in contrast to those of \citet{Poggio2020}, where the only allowed free parameter is the precession rate of the warp. One finds the best fit to the trend of vertical velocity with Galactocentric radius, as derived via a Markov chain Monte Carlo method (MCMC), is described by:
$$
\begin{cases}
\omega_p=-13.57^{+0.20}_{-0.18}\ \text{km}\ \text{s}^{-1}\ \text{kpc}^{-1} \\
R_1=8.87^{+0.08}_{-0.09}\ \text{kpc} \\
R_2=17.78^{+1.56}_{-1.86}\ \text{kpc} \\
R_w=1.20^{+0.28}_{-0.26}\ \text{kpc}^{2-\alpha}\\
\alpha=1.53^{+0.10}_{-0.09}\\
\end{cases}
$$
\begin{figure}
    \centering
    \includegraphics[width=\columnwidth]{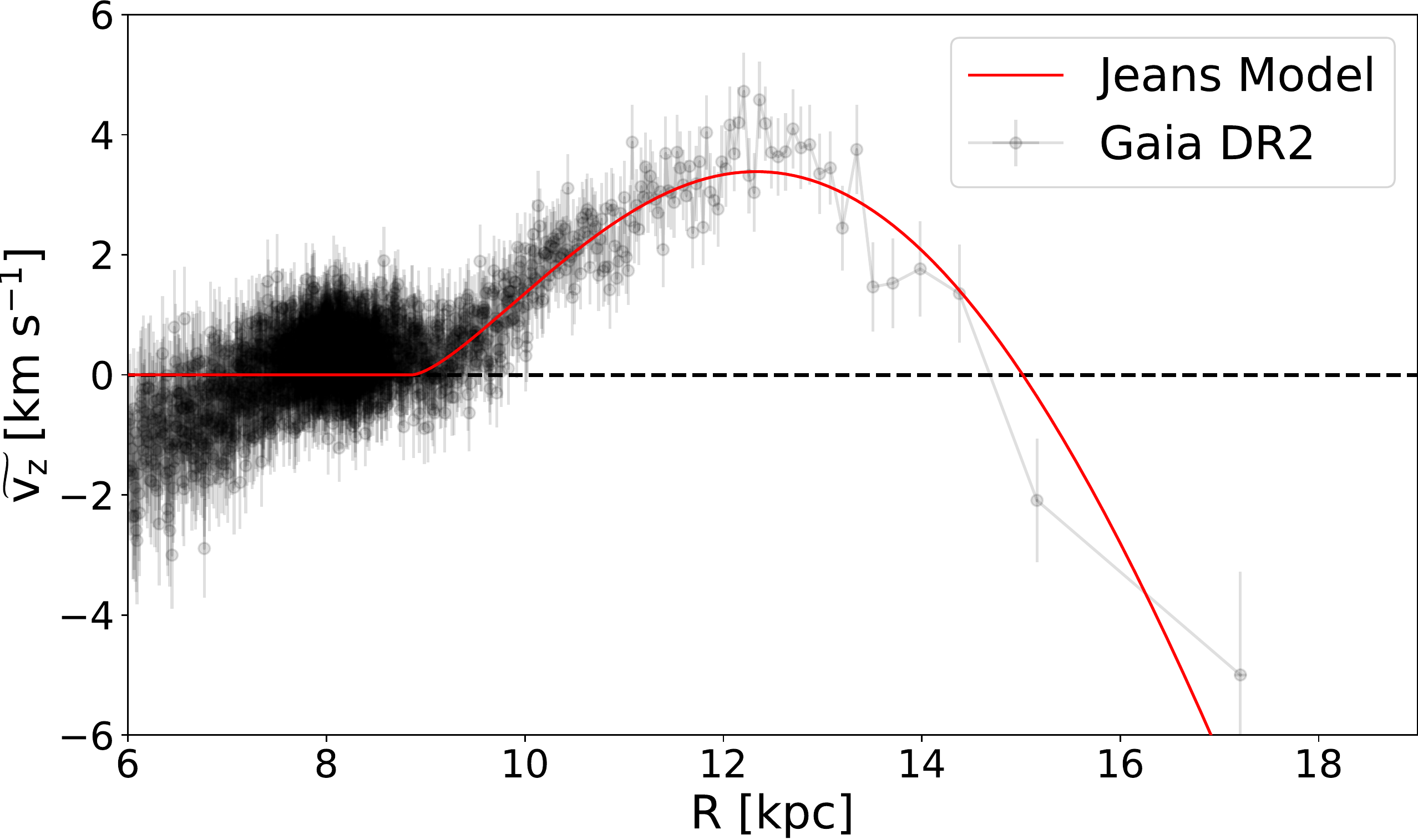}
    \caption{Best fit of our model (red line), inspired by that of \citet{Drimmel2000}, to the {\it Gaia} DR2 data. The model does not work well inside the solar circle because the model is designed for, and constrained by, the outer disk. While we are only fitting $R$ $>$ 8 kpc, the radial range $6<R<8$ is shown because there are claims that the warp starting radius is inside $R=8$ kpc.}\label{fig:jeans_fit}
\end{figure}

\begin{figure}
    \centering
    \includegraphics[width=\columnwidth]{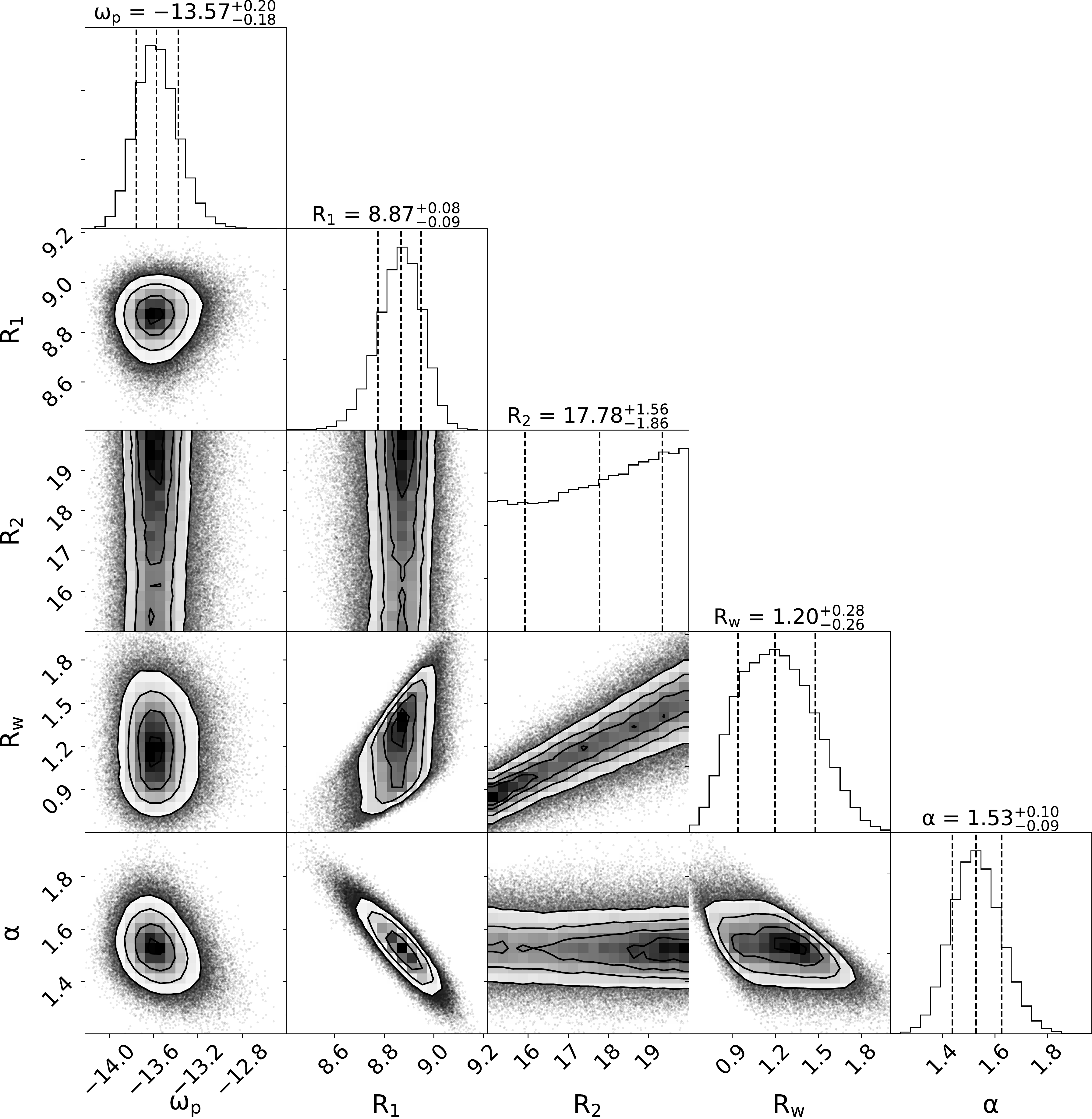}
    \caption{Corner plot of the MCMC fitting of the model. The fact that $R_2$ is not well constrained can be explained by the low number of stars beyond $R>16$ kpc.}\label{fig:jeans_mcmc}
\end{figure}

The best fit is shown in \Cref{fig:jeans_fit} in red, and the corner plot for MCMC fitting is shown in \Cref{fig:jeans_mcmc}. Notice that the model does not match well in the inner part of the Galaxy. This is expected for several reasons:  First, we are only fitting our model for R $>$ 8 kpc. Moreover, the Galactic warp would have very limited effect in the inner (more massive) part of the Galaxy, rendering our model inappropriate there. \Cref{fig:jeans_mcmc} also shows that the model is not sensitive to the ending radius of the warp $R_2$. We attribute that insensitivity to the low number of stars beyond $R>16$ kpc.

Our result for $\omega_p$ is consistent with that of \citet{Poggio2020} (again, for them, $\omega_p$ is the only free parameter and we adopt an opposite sign convention for the direction of the precession term). While our model is very crude in construction, it illustrates the possibility of explaining the decline in vertical velocity as due to a warp precessing in the direction of Galactic rotation.

Even though such a decline was also observed by \citet{Drimmel2000}, they attributed it to the extremely large uncertainty in distance for stars beyond the solar neighborhood. However, that does not appear to be the case here as the number of stars within each of our binned data points is significantly higher, which greatly reduces the uncertainty of the mean value. 

\begin{figure}
    \centering
    \includegraphics[width=\columnwidth]{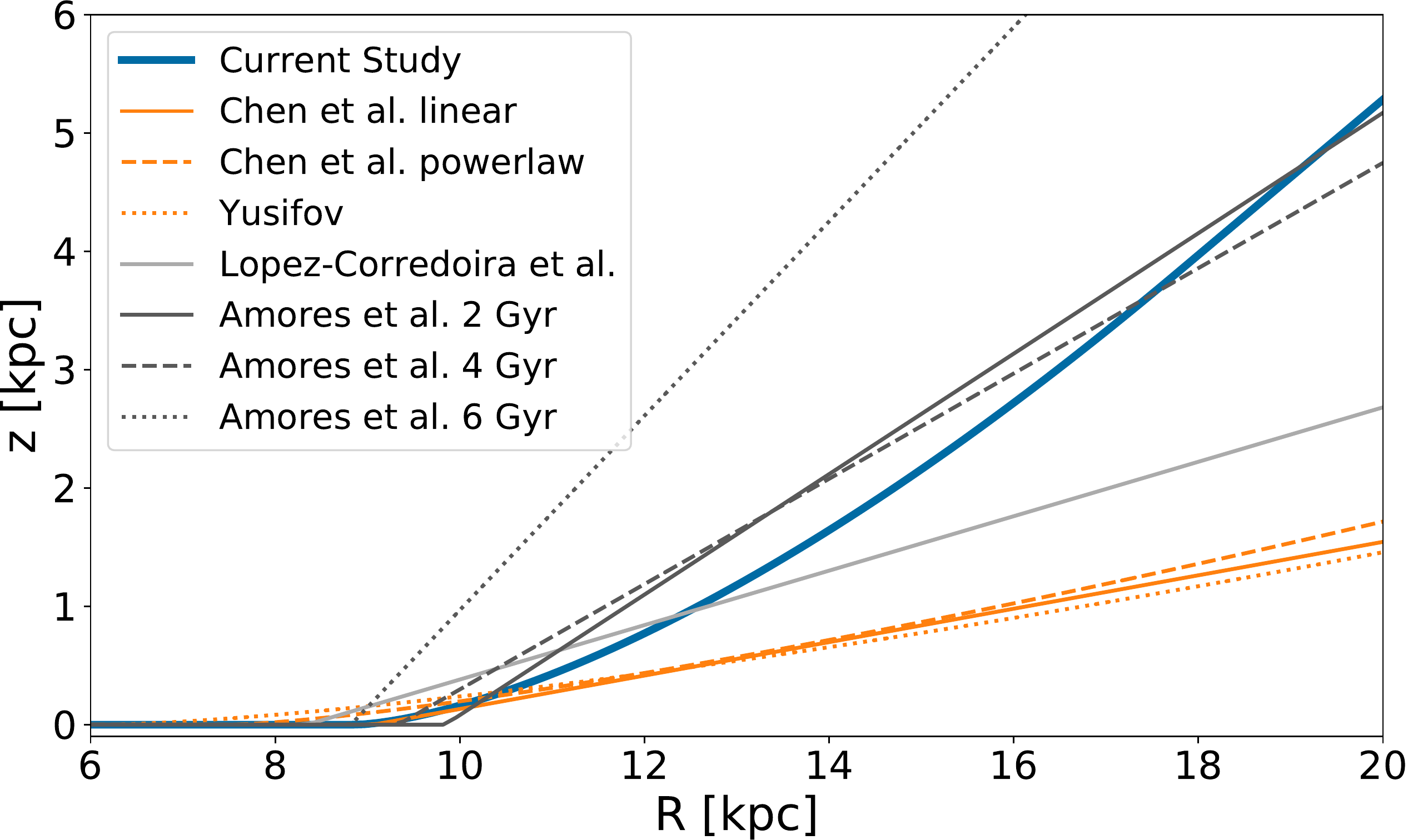}
    \caption{Comparison of existing Galactic warp models by \citet[][their linear and power law models]{Chen2019}, \citet{Yusifov2004}, \citet{Lopez2014}, and \citet[][for three different ages]{Amores2017}. The comparison illustrates that there is large spread in the amplitude of the warp in existing models.}\label{fig:model_comp}
\end{figure}

Our result for the Galactocentric radius where the warp begins agrees well with previously reported values. A comparison of existing models, with trends in $z$ with $R$ shown at the maximum vertical distance from the Galactic midplane for each, is provided in \Cref{fig:model_comp}. However, while the latter figure shows that there is good agreement on $R_1$, there is also a large spread in the amplitude of the warp among the various models.  Moreover, our model agrees better with those exhibiting a stronger warp. It is worth noting that the set of models by \citet{Amores2017}, to which we show the most agreement, have included more physics (e.g., flaring, disk truncation, star formation history, etc.) 
than the other models, including ours, which is a reassuring check on our model.

\section{Age Variations in the Character of the Galactic Warp}\label{sec:age}
In the past few years there have been several lines of evidence suggesting that the parameters of the warp in the Milky Way disk change with the average age of the tracing stellar population \citep[e.g.,][]{Drimmel2000,Amores2017,Romero-Gomez2019,Poggio2020}. In this section we use the stellar age catalog provided by \citet{Sanders2018} to explore how different aged populations are warped differently. This catalog contains the ages of $\sim$3 million \gaia stars, derived using a Bayesian framework to characterise the probability density functions of age for giant stars with combined photometric, spectroscopic, and astrometric information, supplemented with spectroscopic masses, where available. We only include stars for which Sanders \& Das set ``flag = 0''; according to these authors, stars were assigned non-zero flags when (a) the isochrone fitting failed completely, (b) the isochrone overlapped with the data at only one point, (c) the spectroscopic or photometric input data are problematic, or (d) for which the derived ages are unreasonably small ($<100$ Myr).

\begin{figure}
    \centering
    \includegraphics[width=\columnwidth]{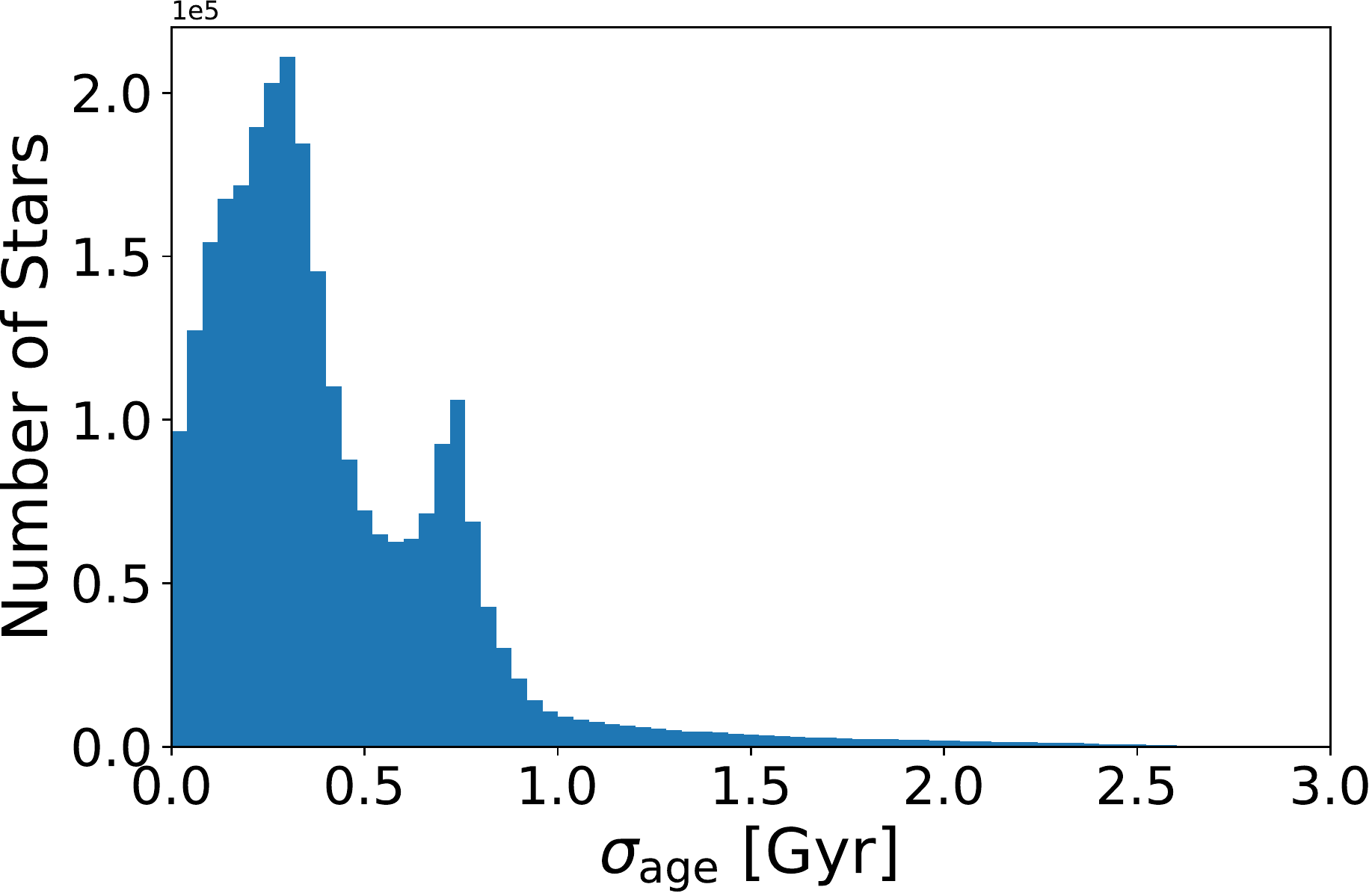}
    \caption{Distribution of error in ages of individual stars in \citet{Sanders2018} catalog.}\label{fig:age_err}
\end{figure}

We acknowledge one caveat is that these ages were derived from extrapolating the relation C/N with age at the solar vicinity. Although individual stars in the \citet{Sanders2018} catalog may have a large uncertainty in their estimated age (see \Cref{fig:age_err}), these estimates are of sufficient quality to sort stars broadly by age and serve as a general indicator of the average age of a population when averaging over a significant number of stars. We selected stars in four age bins: those stars with ages $0-3\ \rm Gyr$ as a ``young population'', those with $3-6\ \rm Gyr$ ages as an  ``intermediate age population'', those $6-9\ \rm Gyr$ in age as an ``old population'', and finally, those dated at $>9\ \rm Gyr$ as an ``ancient population''.  

\begin{figure}
    \centering
    \includegraphics[width=\columnwidth]{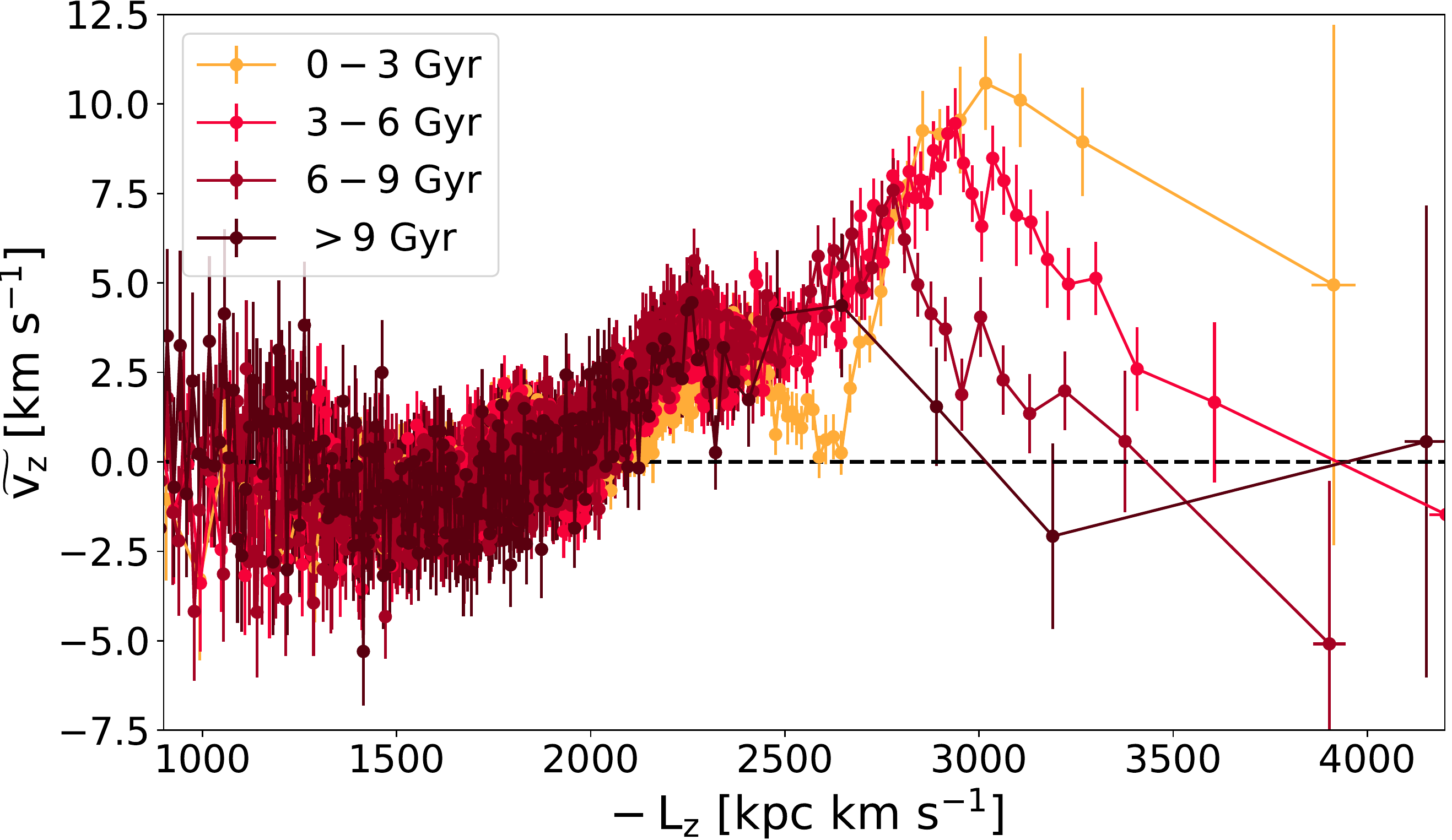}
    \caption{Changes in median $v_z$ as a function of angular momentum with respect to stellar populations of different ages. Note that the young population displays a much larger vertical velocity than the old population, and all of the populations display a downward trend when the angular momentum is large enough.}\label{fig:age}
\end{figure}

\begin{figure*}
    \centering
    \subfloat[0-3 Gyr]{
        \includegraphics[width=\columnwidth]{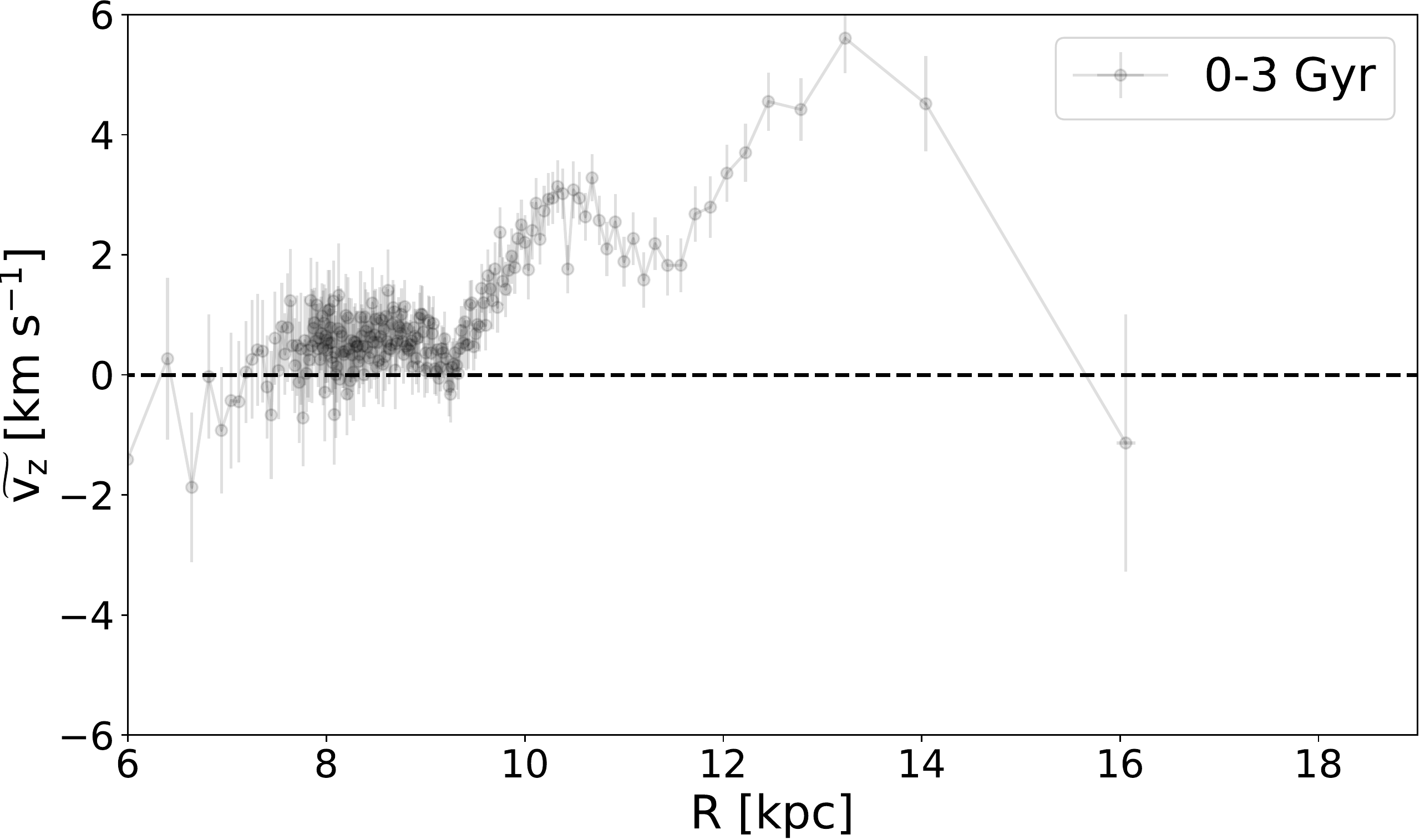}
    }
    \subfloat[3-6 Gyr]{
        \includegraphics[width=\columnwidth]{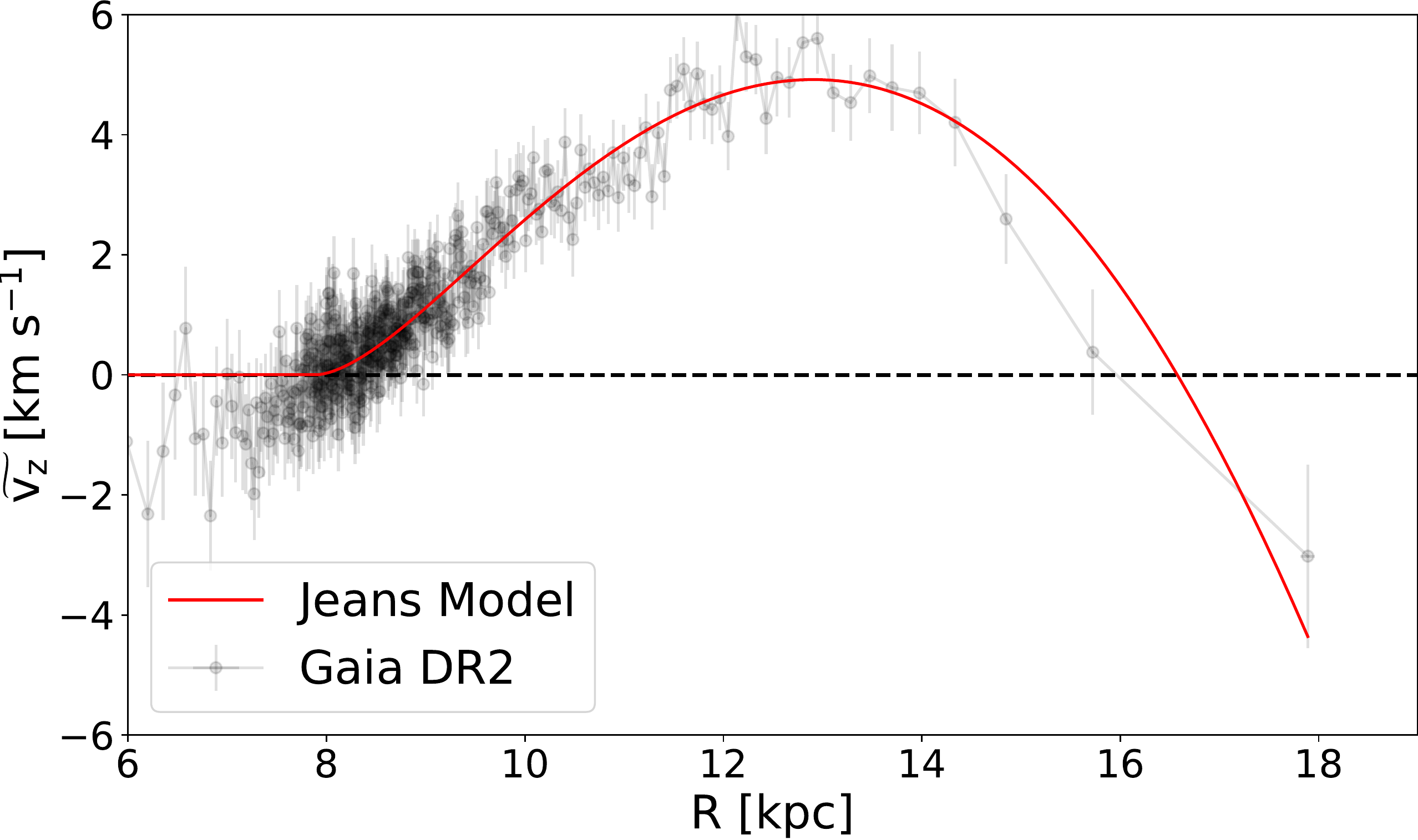}
    }\\ [2ex]
    \subfloat[6-9 Gyr]{
        \includegraphics[width=\columnwidth]{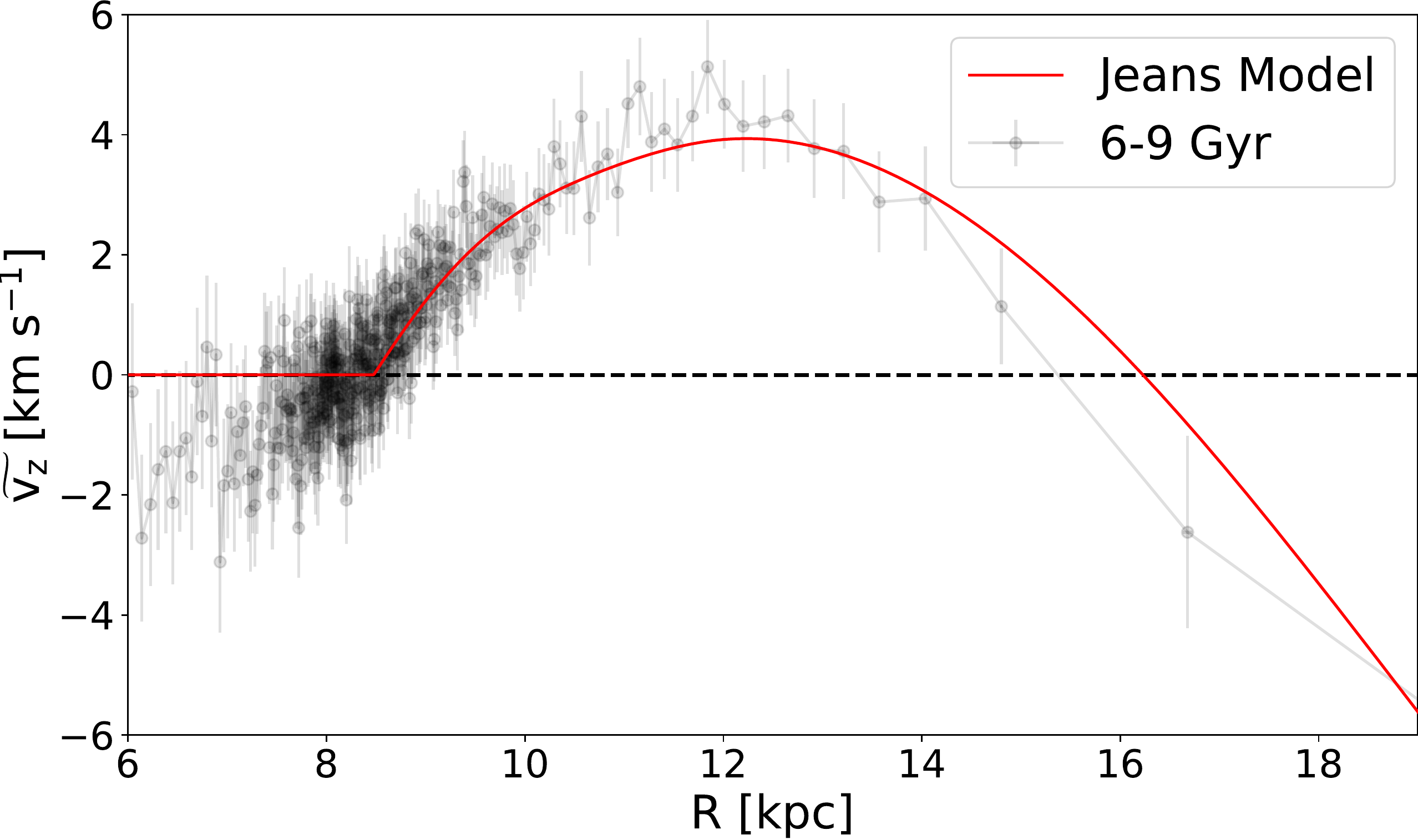}
    }
    \subfloat[$>9$ Gyr]{
        \includegraphics[width=\columnwidth]{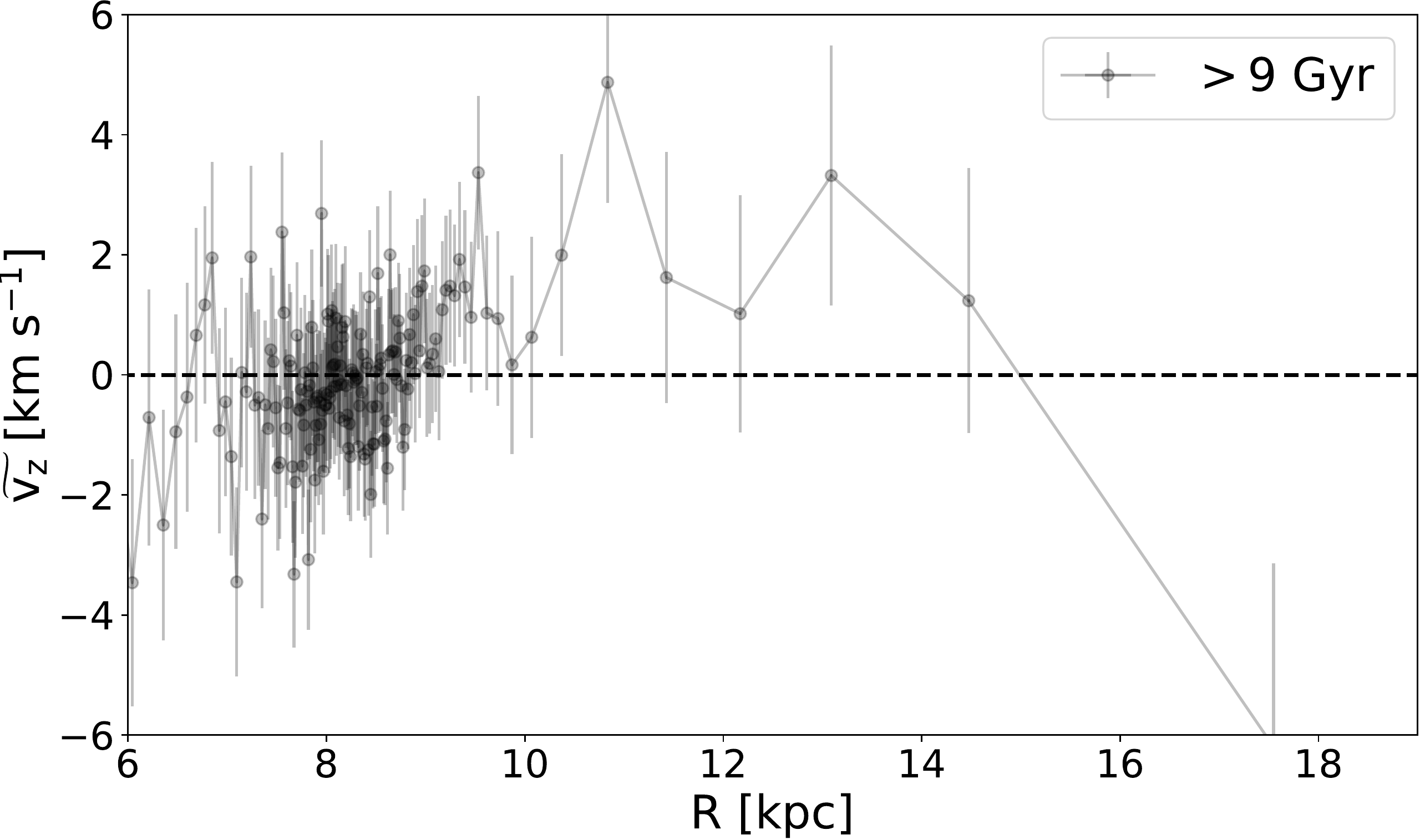}
    }
    \caption{Our simple model fitted to the four different age populations. For the population with stellar ages between 0 and 3 Gyr, a number of ripples between 8 to 14 kpc are detected and the drop-off in velocity is not prominent when compared to the ripples. Our simple model is not complex enough to account for these features. For the population with $>9$ Gyr, due to the large error bars, it is not possible to detect any warp signature, but, on the other hand, the presence of the signature cannot be excluded. We examined the azimuthal velocity of the population and found out that it drops to $\rm <150\ \text{km}\ \text{s}^{-1}$, which indicates a large fraction of stars are from the halo and renders our model inapplicable. Therefore, no fitting is done for the $>9$ Gyr population.}\label{fig:age_pop}
\end{figure*}

\begin{figure}
    \centering
    \includegraphics[width=\columnwidth]{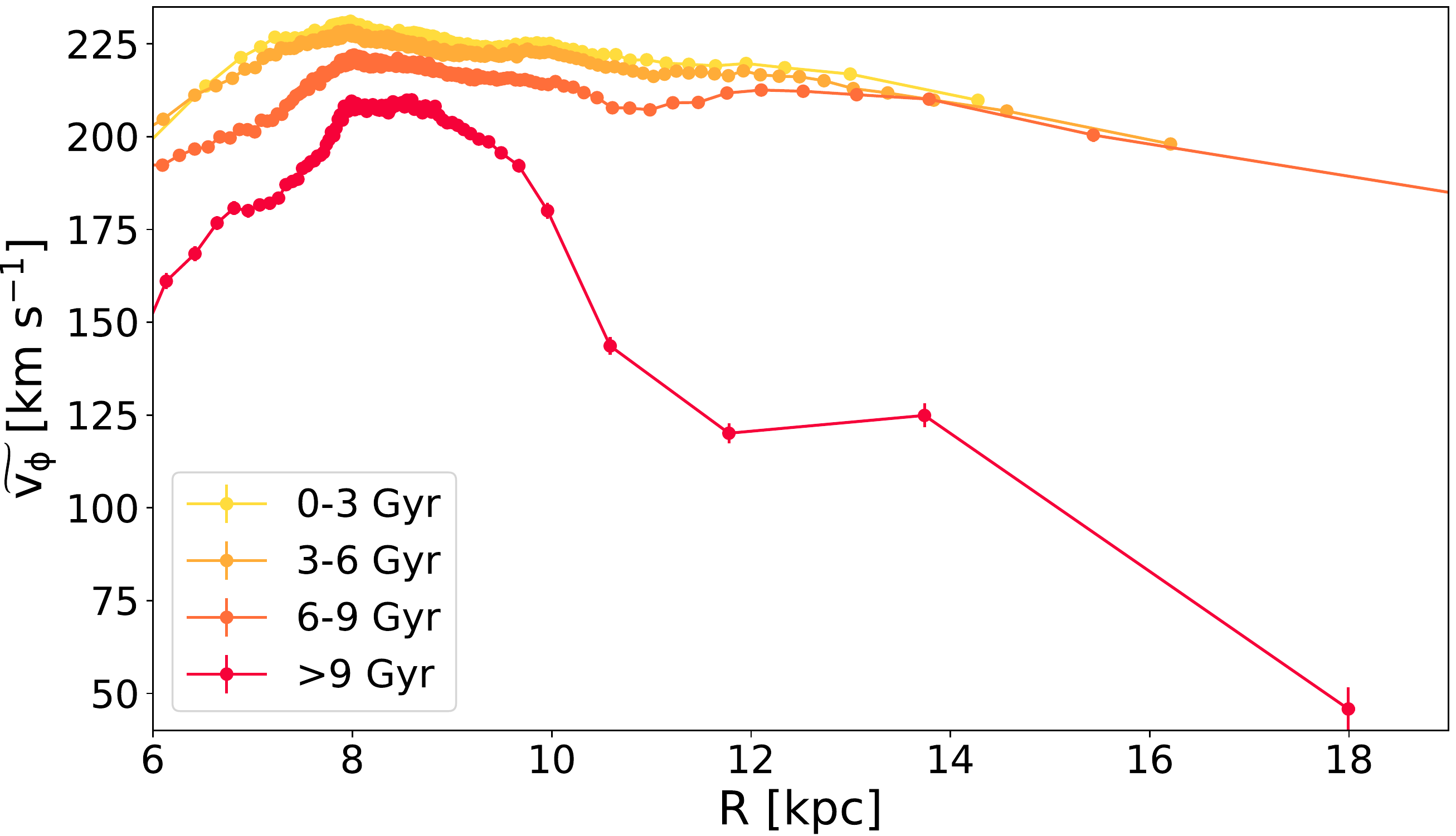}
    \caption{Median azimuthal velocity versus Galactocentric radius for all four populations.  The rotation curve is no longer flat in the outer part of the Galaxy for the ancient population, which indicates that this population is likely dominated by halo stars in the outer part.}\label{fig:vphi_9}
\end{figure}

The mean vertical velocity versus angular momentum for each of these age groups is shown in \Cref{fig:age}. It is clear that there are major differences in this particular kinematical trend between different age populations.  The young population (orange points in \Cref{fig:age}) shows the largest increase in vertical velocity, and the maximum median $v_z$ declines with increasing population age through the intermediate and old aged populations. The abrupt decline in median $v_z$ is evident in all three populations with age $<$9 Gyr, albeit with slightly differing starting $L_z$ for the beginning of the drop-off. For the ancient stars (brown points in \Cref{fig:age}) the effect of the warp is less evident; this is likely due to the large number of halo stars within the ancient population.  This conclusion is based on the character of the rotation curve exhibited by this population, which, unlike the younger star groups, shows a rapid decline beyond the solar circle \Cref{fig:vphi_9}.

We applied our simple analytical model to fit and track the changes of parameters of the warp with stellar age in \Cref{fig:age_pop}. However, because our model is limited in its complexity, it cannot account for ripples not associated with the Galactic warp or non-thin-disk stellar kinematics. As a result, fitting results are not reported for the youngest population, for which prominent substructures not related to the Galactic warp are attributable to the higher frequency ripples. Nor do we report a fit for the ancient population, where, as we have shown \Cref{fig:vphi_9}, a substantial fraction of the sample is contaminated by halo stars. 

On the other hand, for the 3-6 Gyr population, our fit yields a precession rate of $-11.59^{+0.30}_{-0.25}\ \text{km}\ \text{s}^{-1}\ \text{kpc}^{-1}$ while for the 6-9 Gyr population we obtain $-12.19^{+0.49}_{-0.39}\ \text{km}\ \text{s}^{-1}\ \text{kpc}^{-1}$.
The lack of any significant difference between these two populations suggests that the response to the warp in at least these two populations is similar. However, from \Cref{fig:age}, a clear difference in the size of the vertical velocity is present between different age populations, with the older population being slower.  This difference in amplitude could be consistent with the warp being a recent event (that is, within the past 3 Gyr), but where different aged populations respond differently in bulk: Presumably the older population, which is also the kinematically hotter population, would have a weaker response to dynamical perturbations.

Apart from differences in the amplitude of the warp in different populations, we also find that the peaks of vertical velocity are at different Galactocentric radius for different age populations (\Cref{fig:age_pop}). The peak vertical velocity is moving closer to the Galactic Center as the population grows older. One explanation for this is suggested by \Cref{fig:vphi_9}, where a decrease in azimuthal velocity correlates to  older populations; according to the factor ($\frac{v_\phi}{R}+\omega_p$) in Equation 9, when the precession rate is similar between two populations, the population with smaller azimuthal velocity will have a peak closer to the Galactic Center. However, we also notice, curiously, that the fractional decrease in azimuthal velocity (that is, from $\sim$220 km s$^{-1}$ for the 3-6 Gyr population to  $\sim$210 km s$^{-1}$ for the 6-9 Gyr population) is about a factor of two smaller than the fractional decrease in where the peak vertical velocity is located ($\sim$13 kpc for the 3-6 Gyr population, $\sim$12 kpc for the 6-9 Gyr population), when these decreases should be proportional.

\begin{figure}
    \centering
    \includegraphics[width=\columnwidth]{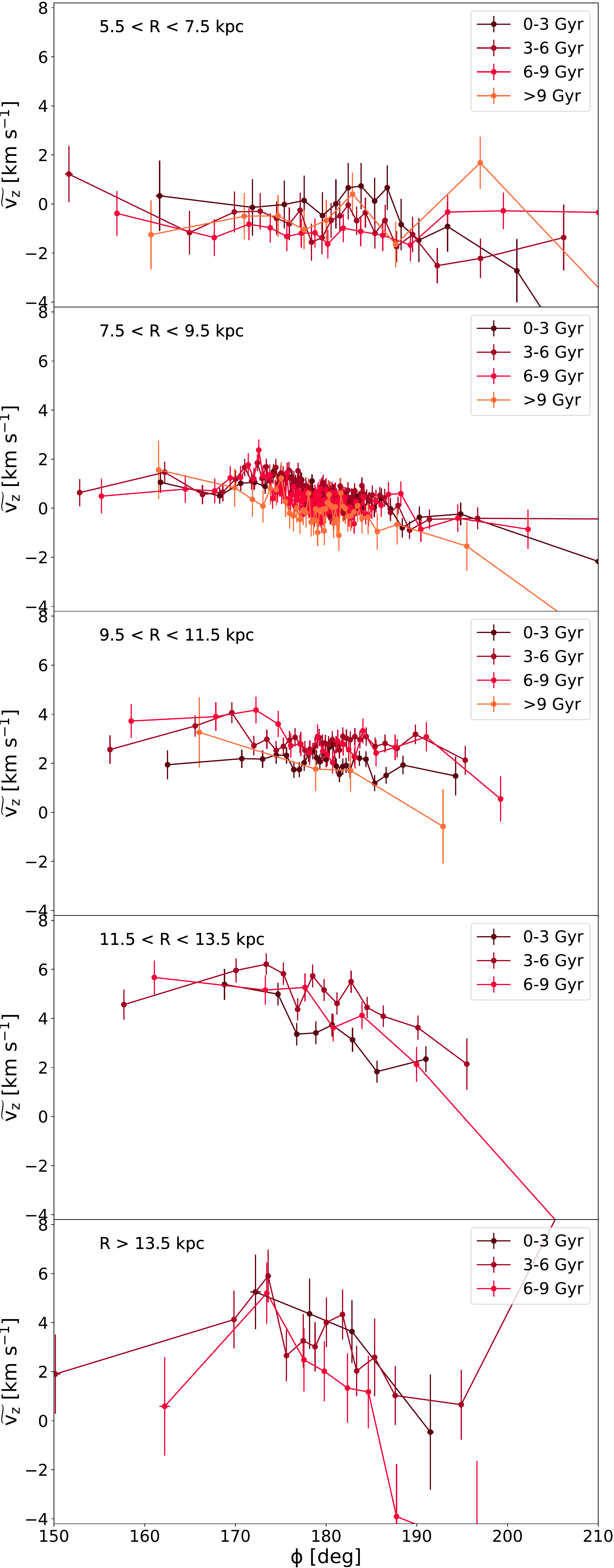}
    \caption{Vertical velocity as a function of Galactocentric azimuthal angle for different age bins. Stars with age $>9$ Gyr and Galactocentric radius $R>11.5$ kpc are not included due to the population being dominated by halo stars.}\label{fig:phi_vz_age}
\end{figure}

With no age-variable signatures in the precession of the warp but some differences in the velocity amplitude, it is worth testing whether there may be age-variable signatures in the {\it lopsidedness} of the warp that we previously found across the entire sample (Section \ref{sec:v_z}). \Cref{fig:phi_vz_age} shows the azimuthal distribution of median vertical velocity in different age populations for different radial annuli. The lopsidedness is prominent in all age groups for radii beyond $R>7.5$ kpc. Moreover, the lopsidedness remains similar, with the vertical velocity increasing when $\phi<180$ deg and decreasing when $\phi>180$ deg. The slope of increase and decrease is also similar across the different age populations. This further supports that the different age population has similar response to the Galactic warp, thus suggesting a possible gravitational origin.

In the end, our consideration of potential age differences in the characteristics of the warp reveals them to be consistent with a model whereby the intermediate and older populations are both responding to a single gravitational perturbation happening less than 3 Gyr ago.

\section{Conclusions}
\label{sec:conclusion}
In this study we combine the precise stellar abundances from the APOGEE survey with the astrometry from \gaia DR2 and the StarHorse distance computed by \citet{Queiroz2020} to study the vertical and radial velocity components of stars with respect to the Galactocentric radius and angular momentum.  We take advantage of the detailed and accurate chemical abundances available in the smaller APOGEE-\gaia sample (\Cref{fig:apogee_selection}) as a guide to interpretation of the much larger {\it Gaia}-only sample. Our analysis probes disk kinematics to a greater Galactocentric radius ($R\sim 18$ kpc) than has been explored previously (\Cref{fig:dist}). From these combined data we find evidence for the Galactic warp and characterize its onset radius and precession rate. Interestingly, a number of high spatial frequency kinematical features are also found, as has been reported by previous authors at smaller Galactocentric radii (\Cref{fig:vz_vs_lz} and \Cref{fig:vr_vs_lz}).

We find that over a large range of $L_z$ the overall median stellar vertical velocity $\widetilde{v_z}$ increases with $L_z$. Moreover, the increase of the mean vertical velocity is more pronounced for $L_z$ $>$ 1800 kpc km s$^{-1}$ and continues until $L_z$ $\sim$ 2800 kpc km s$^{-1}$ or $R=13\ \text{kpc}$, after which the vertical velocity sharply declines (\Cref{fig:vz_vs_lz} and \Cref{fig:vz_vs_r}). This abrupt decrease in $\widetilde{v_z}$ is reported for the first time. We associate this entire global trend in $\widetilde{v_z}$ as a signature of the Galactic warp.  We also study the vertical velocity as a function of the Galactocentric azimuthal angle for the \gaia sample, and found differences in this parameter with respect to the Galactocentric azimuthal angle for $\phi$ $<$ $180^{\circ}$ and $\phi$ $>$ $180^{\circ}$, evidence consistent with a warp line-of-nodes toward this anticenter direction (\Cref{fig:phi_vz}). However, the velocity trends with $\phi$ in our data appear to be asymmetric about $\phi \sim 180^{\circ}$ (\Cref{fig:phi_vz_absdel}), which is evidence suggesting that the Galactic warp may be lopsided.

An analytical model using the Jeans Equation with consideration of a non-zero radial motion is constructed to explain the observed phenomena, and shows that the declining trend in vertical velocity can be explained as a manifestation of the Galactic warp. We find that the warp has a starting radius of $8.87^{+0.08}_{-0.09} \text{kpc}$ and a precession rate of $-13.57^{+0.20}_{-0.18}\ \text{km}\ \text{s}^{-1}\ \text{kpc}^{-1}$ (\Cref{fig:jeans_fit} and \Cref{fig:jeans_mcmc}), a value slightly higher than the 10.86\ \text{km}\ \text{s}$^{-1}\ \text{kpc}^{-1}$ reported recently in \cite{Poggio2020} (accounting for the  opposite  sign  convention  we adopt for  the  direction  of the precession term compared to \citealt{Poggio2020}). Note that the parameters related to the warp itself, namely the Galactocentric radius where the warp starts and ends ($R1$ and $R2$, respectively), the scale height of the warp ($r_{h}$), and the precession speed of the warp ($\omega_{p}$) are free parameters in our fitting procedure, whereas \cite{Poggio2020} only allowed as a free parameter the precession rate of the warp. Furthermore, our model illustrates that the reported decline in vertical velocity can be explained due to a warp precessing in the direction of the Galactic rotation.

We compare the spatial amplitude of our model with those of other existing models, for which there is a large spread in values (\Cref{fig:model_comp}). Our model agrees better with others exhibiting a stronger warp, with best match to those by \citet{Amores2017}, for which markedly additional physics is considered (e.g., flaring, disk truncation, star formation history, etc.) than is typical for other studies, including our own.

Using two stellar populations of different ages, young (OB-type) stars and intermediate-old age (red giant branch, RGB) stars, several authors have reported that the parameters of the warp in the Milky Way disk change with the average age of the tracing stellar population \citep[e.g.,][]{Drimmel2000,Romero-Gomez2019,Poggio2020}.  Here we used the stellar age catalog provided by \cite{Sanders2018} to explore how different aged populations are warped differently.  We find that different aged populations show similar warp characteristics, except for velocity amplitude. The young population (0-3 Gyr) shows the largest increase in vertical velocity, and the maximum median $v_{z}$ declines with increasing population age through intermediate (3-6 Gyr) and old (6-9 Gyr) populations (\Cref{fig:age}). We also find that the abrupt decline in median $v_{z}$ is present in all three populations with age $<$9 Gyr, albeit with slightly differing starting $L_{z}$ for the beginning of the drop-off. The effect of the warp for the ancient stars ($>$9 Gyr) is less evident; this is likely due to the large number of halo stars within the ancient population (\Cref{fig:vphi_9}). 

We also applied our simple analytical model to track the changes of other warp parameters with stellar age. For example, for the 3-6 Gyr population our model fit yields a precession rate of $-11.59^{+0.30}_{-0.25}\ \text{km}\ \text{s}^{-1}\ \text{kpc}^{-1}$, while for the 6-9 Gyr population we obtain $-12.19^{+0.49}_{-0.39}\ \text{km}\ \text{s}^{-1}\ \text{kpc}^{-1}$ (\Cref{fig:age_pop}). Meanwhile, the vertical velocity as a function of Galactocentric azimuthal angle for different age populations and radial annuli shows that the lopsidedness remains similar for these two populations (\Cref{fig:phi_vz_age}). 

Taken together, our study of the warp characteristics with stellar age shows similarities (precession rate and lopsidedness) and differences (velocity amplitude) that are consistent with a scenario where the Galactic warp seen in 3-9 Gyr aged stars reflects their response to a more recent ($<$3 Gyr) gravitational interaction, for example a perturbation in the disk incited by a satellite galaxy. 

\acknowledgments
We thank the anonymous referee for a thorough review of the paper that both led to the correction of an oversight in our original draft and improved the overall quality of the presentation.

This work has made use of data from the European Space Agency (ESA) mission {\it Gaia} (\url{https://www.cosmos.esa.int/gaia}), processed by the {\it Gaia} Data Processing and Analysis Consortium (DPAC, \url{https://www.cosmos.esa.int/web/gaia/dpac/consortium}). Funding for the DPAC has been provided by national institutions, in particular the institutions participating in the {\it Gaia} Multilateral Agreement.

Funding for the Sloan Digital Sky Survey IV has been provided by the Alfred P. Sloan Foundation, the U.S. Department of Energy Office of Science, and the Participating Institutions. SDSS acknowledges support and resources from the Center for High-Performance Computing at the University of Utah. The SDSS web site is www.sdss.org.

SDSS is managed by the Astrophysical Research Consortium for the Participating Institutions of the SDSS Collaboration including the Brazilian Participation Group, the Carnegie Institution for Science, Carnegie Mellon University, the Chilean Participation Group, the French Participation Group, Harvard-Smithsonian Center for Astrophysics, Instituto de Astrofísica de Canarias, The Johns Hopkins University, Kavli Institute for the Physics and Mathematics of the Universe (IPMU) / University of Tokyo, the Korean Participation Group, Lawrence Berkeley National Laboratory, Leibniz Institut für Astrophysik Potsdam (AIP), Max-Planck-Institut für Astronomie (MPIA Heidelberg), Max-Planck-Institut für Astrophysik (MPA Garching), Max-Planck-Institut für Extraterrestrische Physik (MPE), National Astronomical Observatories of China, New Mexico State University, New York University, University of Notre Dame, Observatório Nacional / MCTI, The Ohio State University, Pennsylvania State University, Shanghai Astronomical Observatory, United Kingdom Participation Group, Universidad Nacional Autónoma de México, University of Arizona, University of Colorado Boulder, University of Oxford, University of Portsmouth, University of Utah, University of Virginia, University of Washington, University of Wisconsin, Vanderbilt University, and Yale University.
\vspace{5mm}
\facilities{}
\software{astropy \citep{Astropy},
          corner.py \citep{corner},
          emcee \citep{emcee},
          topcat \citep{topcat}}

\bibliography{sample63}{}
\bibliographystyle{aasjournal}

\appendix
\section{Modeling the Effect of Distance Uncertainties and Small Sample Size on the Vertical Motion}

To examine the effect of distance uncertainties and low star sample size on the vertical motion observed, we made use of the Gaia DR2 mock catalog from \citet{Rybizki2018} and applied the same selection criteria as described in Section \ref{sec:data}. The mock catalog was divided into bins of size 40,000 stars, and the uncertainty in distance for stars in the mock catalog were estimated from counterpart stars in the observed catalog in the same distance bin.  The uncertainties for each star in the observed catalog were calculated from the Starhorse \citep{Queiroz2018} distance distribution parameters for the 84th percentile and the 16th percentile, which was adopted as $2\sigma$.
The error distribution of the distances were assumed to have a log-normal distribution. 

\begin{figure}
    \centering
    \includegraphics[width=0.9\columnwidth]{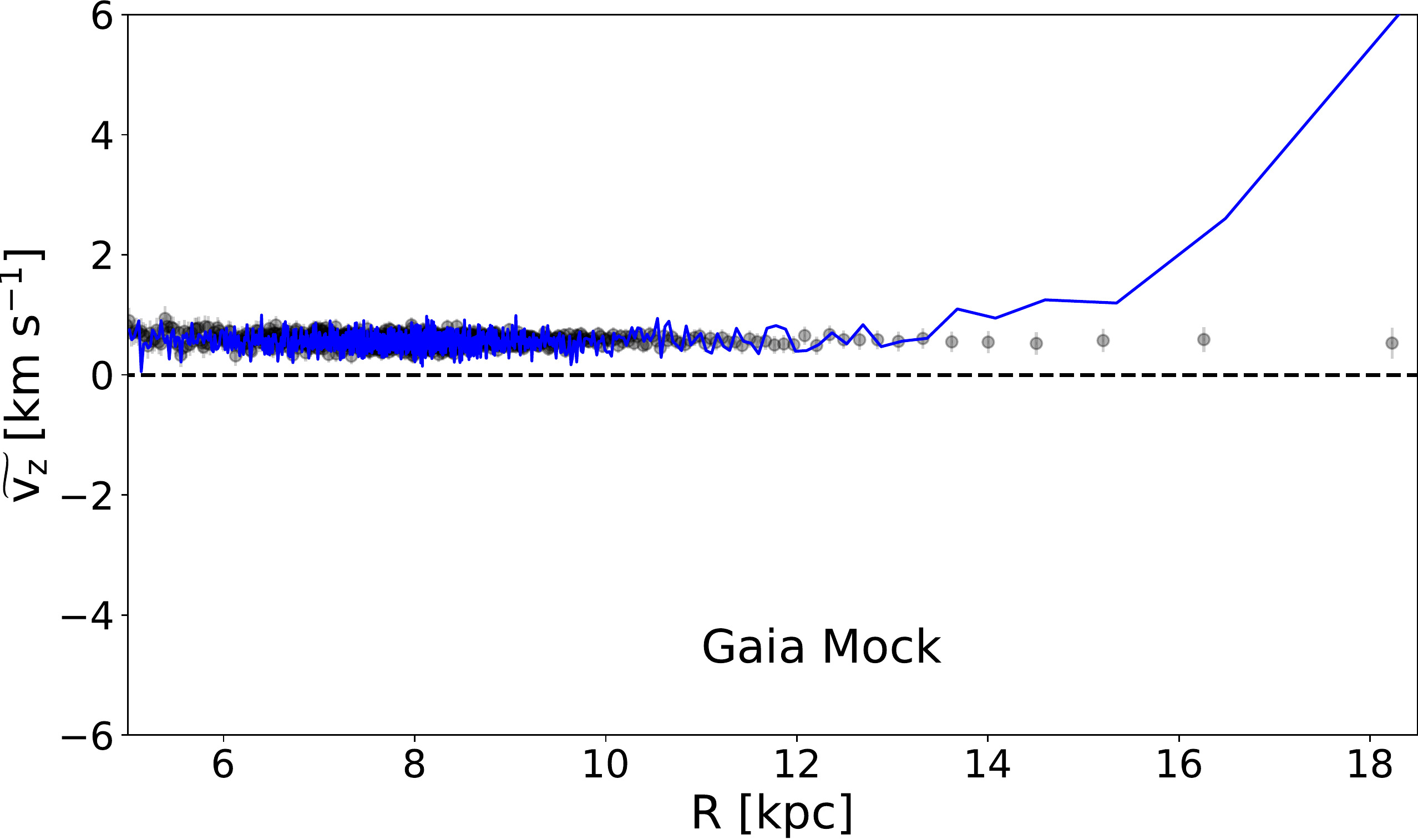}
    \caption{The real and observed vertical velocity from Gaia DR2 Mock catalog. The real vertical velocity is shown with grey circles and the observed vertical velocity (with non-Gaussian distance uncertainties) is represented by the blue line.}\label{fig:d_uncertainty}
\end{figure}

With this simple model, we find that the effect on the observed vertical motion is negligible when $R < 12\ \text{kpc}$. The offset reaches $2\ \text{km}\ \text{s}^{-1}$ at  $R \sim 16\ \text{kpc}$, and then becomes significant beyond that (Fig. \ref{fig:d_uncertainty}). Since the vertical velocity reaches its peak at $R\sim13\ \text{kpc}$ (\Cref{fig:vz_vs_r}), where the effect is still small compared to the changes in vertical velocity, we conclude that the observed decrease in vertical velocity does not come from either distance uncertainties or low numbers of stars in the outer part of the Galaxy. This mock catalog test also shows that the observed velocity will be larger than the real velocity, as shown in \Cref{fig:d_uncertainty}. Thus, it is possible that the warp is either precessing even faster or has an even larger amplitude.

\end{document}